\documentclass[aps,prb,twocolumn,citeautoscript]{revtex4-1}        
\usepackage{amsmath,amssymb,mathrsfs,bm,feynmf,setspace}
\usepackage{graphicx}
\usepackage[tight]{subfigure}    
\usepackage{color} 
\usepackage[colorlinks=true]{hyperref} 
\hypersetup{
    bookmarks=true,         
    unicode=false,          
    pdftoolbar=true,        
    pdfmenubar=true,        
    pdffitwindow=false,     
    pdfstartview={FitH},    
    pdftitle={My title},    
    pdfauthor={Author},     
    pdfsubject={Subject},   
    pdfcreator={Creator},   
    pdfproducer={Producer}, 
    pdfkeywords={keyword1} {key2} {key3}, 
    pdfnewwindow=true,      
    colorlinks=true,       
    linkcolor=magenta, 
    citecolor=blue,        
    filecolor=magenta,      
    urlcolor=cyan           
} 
\newcommand{\al}{\alpha}  
 
\newcommand{\g}{\gamma}
\newcommand{\de}{\delta}

\newcommand{\ka}{\kappa}
\newcommand{\la}{\lambda}

\newcommand{\s}{\sigma}

\newcommand{\w}{\omega}

\newcommand{\pd}{\partial}

\newcommand{\beq}{\begin{equation}}
\newcommand{\eeq}{\end{equation}}
\newcommand{\Beq}{\begin{eqnarray}}
\newcommand{\Eeq}{\end{eqnarray}}
\newcommand{\bml}{\begin{multline}}

\newcommand{\eeqm}{\end{multline}}

\newcommand{\bsp}{\begin{split}}
\newcommand{\esp}{\end{split}}

\renewcommand{\b}[1]{{\bm #1}}
\renewcommand{\t}{\tilde}

\newcommand{\inv}{^{-1}}

\newcommand{\mc}{\mathcal}

\renewcommand{\t}{\tilde}
\newcommand{\ra}{\rightarrow}
\newcommand{\req}[1]{Eq.~(\ref{eq:#1})}

\newcommand{\rfig}[1]{Fig.~\ref{fig:#1}}
\newcommand{\nn}{\nonumber}

\DeclareMathOperator{\tr}{tr}
\DeclareMathOperator{\diag}{diag}
\DeclareMathOperator{\sgn}{sgn}
\DeclareMathOperator{\re}{Re} 
\DeclareMathOperator{\im}{Im}

\newcommand{\si}{\;\;\,}  

\begin{document}

\title{Quantum critical charge response from higher derivatives: \\ is more different?}          
 \author{William Witczak-Krempa}
 \affiliation{Perimeter Institute for Theoretical Physics, Waterloo, Ontario N2L 2Y5, Canada}
  \date{\today}
\begin{abstract}  
We present new possibilities for the charge response in the quantum critical regime in 2+1D using
holography, and compare them with field theory and recent quantum Monte Carlo results. 
We show that a family of (infinitely many) higher derivative terms in the gravitational bulk leads to behavior far richer than what was previously obtained. For example, we prove that the conductivity  
becomes unbounded, undermining previously obtained constraints. We further find
a non-trivial and infinite set of theories that have a self-dual conductivity. Particle-vortex
or S duality plays a key role; notably, it maps theories with a finite number of bulk  
terms to ones with an infinite number. Many properties such as sum rules and stability 
conditions are proved.  
\end{abstract} 
\maketitle 
\singlespacing 
\tableofcontents   

\section{Introduction} 
Continuous quantum phase transitions often lead to the emergence of non-trivial 
degrees of freedom from simple ones\cite{keimer,*deconfined,*ssl_tasi}. The former can result from a high level of entanglement between the latter,  
requiring 
a non-quasiparticle description. 
One example comes from the Mott transition of bosons at integer filling in 2+1D, which can be described by a
strongly interacting conformal field theory (CFT) at low energies. A natural and experimentally relevant knob
is temperature, and it remains a challenge to understand how the continuum 
of quantum critical excitations responds to it. Valuable lessons can be learned by 
studying more accessible limits of the problem. The holographic AdS/CFT correspondence\cite{Maldacena} offers one such 
avenue by mapping certain CFTs to higher dimensional classical gravity.  
Focusing on correlations of conserved conserved currents, more precisely the frequency-dependent charge conductivity $\s(\w/T)$, 
we note that interesting properties have been predicted using holography: duality relations\cite{m2cft}, bounds\cite{buchel,ritz,myers11,hofman09,hofman},  
sum rules\cite{sum-rules,ws,ws2}, existence of special damped excitations\cite{star1,ws,ws2}, etc.  
As these were derived using limited holographic actions,
it is natural to question their robustness to higher derivative (HD) terms. This becomes especially
important when comparing with CFTs of relevance to condensed matter\cite{ads-qmc}.  
\begin{figure}
\centering%
\includegraphics[scale=.43]{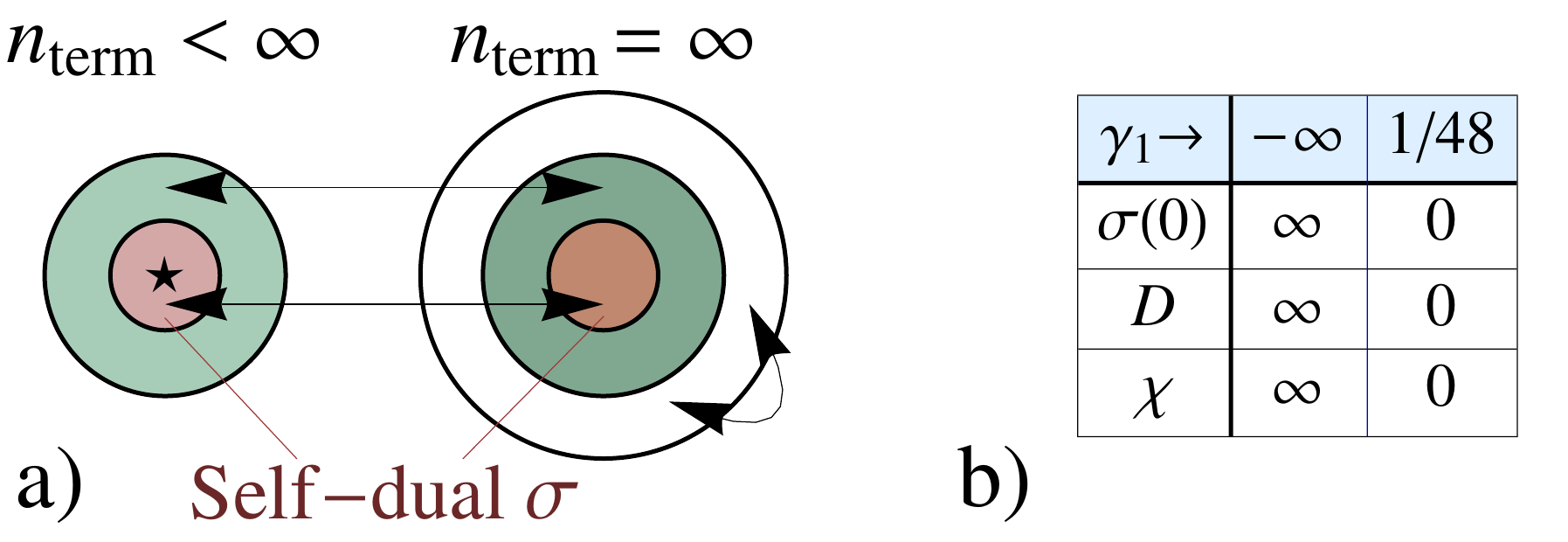}      
\caption{\label{fig:map} 
a) Sketch of the subspace of holographic theories describing the charge 
response of CFTs in 2+1D. It encompasses any action with $n_{\rm term}$ terms containing 2 field strengths   
coupled to curvature tensors. The arrows indicate S-duality. The central regions (red) have a self-dual, $\w$-independent 
conductivity $\s$. The star corresponds to the 
R-charge response of superconformal Yang-Mills. b) New allowed asymptotics for the d.c.\ conductivity, diffusion 
constant and susceptibility; $\g_1$ parameterizes a 6-derivative term.  
}  
\end{figure}

We explore a large subspace of allowed HD terms in the holographic bulk and find 
qualitatively new physics. For instance, previously-derived bounds\cite{ritz,myers11} for $\s(\omega/T)$ become strongly violated;
$\s$ can even approach the conductivity of O($N$) CFTs at large-$N$, displaying an arbitrarily 
sharp Drude-like peak. We have also discovered an infinite and non-trivial family of theories with a  
$\w$-independent, self-dual $\s$. A key concept in the analysis is that of generalized 
particle-vortex (S) duality\cite{witten}, 
which we show maps theories with a finite number of derivatives to ones with an infinite number. 
\rfig{map} sketches the space of theories we study and their relations. 

\section{Framework} 
The gauge/gravity correspondence maps a conserved U(1) current in the boundary CFT 
to a U(1) gauge field in the gravitational bulk. 
One needs to identify a proper action for
the metric and gauge field, with the understanding that the latter is treated in the probe limit,
as relevant for linear response. For a special CFT\cite{abjm}, namely the $\mc N=8$ superconformal fixed point of Yang-Mills in the
large-$N$ limit in 2+1D, the action is Einstein-Maxwell in the
presence of a negative cosmological constant\cite{m2cft}. The gauge field thus evolves according to Maxwell's
equations in the curved spacetime with metric: $ds^2=u^{-2}[-f(u)dt^2+dx^2+dy^2+L^2 du^2/f(u)]$, where
$f(u)=1-u^3$. This spacetime is asymptotically anti de Sitter, with cosmological constant $\propto -1/L^2$,
and contains a black hole with a horizon at $u=1$. It describes the thermally excited CFT, which
can be thought of living at the boundary of the spacetime, $u=0$. Since we are interested in applying
the holographic principle to CFTs other than super-Yang-Mills, we adopt an effective-field-theory spirit by
adding symmetry-allowed terms with more derivatives. In principle, these can be generated 
by allowing for departures from the limit of infinite 't Hooft coupling, which controls the string tension
in the holographic bulk.  

Let us begin with the action (setting $L=1$)
\begin{multline}\label{eq:S}
  S=\int_x \Big[\frac{1}{2\ka^2}(R+6)+g_4^{-2}\big(-\tfrac{1}{4}F^2+\g C_{abcd}F^{ab}F^{cd}\\   
 +\g_1 C^2F^2+\g_2 C_{abef}C^{efcd}F^{ab}F_{cd}+\cdots\big)\Big]\,,
\end{multline} 
where $\int_x=\int d^4x\sqrt{-g}$, $C_{abcd}$ is the Weyl curvature tensor 
(traceless part of the Riemann tensor, $R_{abcd}$),
and $C^2=C_{abcd}C^{cdab}$. $g_4$ is the bulk gauge coupling, it is associated with the $T=0$ conductivity $\s(\w/T=\infty)=g_4^{-2}$. 
Restricting oneself to terms with   
4 derivatives and less, the only necessary addition to Einstein-Maxwell in order
to study linear response in the charge sector is the one parameterized by $\g$ in \req{S}\cite{myers11}.  
The new terms, parameterized by $\g_{i=1,2}$ etc, contain 6 or more derivatives and thus go 
beyond the Weyl action.  
In this paper, we restrict ourselves to terms which can  
be constructed out of 2 field strengths\footnote{This is sufficient for linear response.} 
and any number of curvature tensors. This 
dispenses with pure-gravitational HD terms such as higher powers of the Ricci scalar, or terms with
covariant derivatives acting on the $F$'s. 
The reasons for
this are manifold. First, it provides a transparent yet versatile framework: the infinite family of terms
gives rise to a rich landscape of CFT correlation functions without the need to solve non-linear
equations for the metric, for e.g.. Second, our subspace of terms is   
closed under electric-magnetic (EM) duality in the bulk, which corresponds to 
S duality in the boundary CFT\cite{witten}.  

In order to simplify the calculations, it will be useful to write down the symmetry-allowed
action for $A_a$ in the general form\cite{myers11}:
\begin{align}
  S_A &=\int_x \frac{-1}{8g_4^2} F_{ab}X^{abcd}F_{cd}\, ; \label{eq:X} \\
 \frac{X_{ab}^{\si cd}}{8} &= \frac{I_{ab}^{\si cd}}{8} -\g C_{ab}^{\si cd} -\g_1 C^2\frac{I_{ab}^{\si cd}}{2}   
  -\g_2 C_{ab}^{\si ef}C_{ef}^{\si cd} -\cdots ,\nn
\end{align}
where $I_{ab}^{\si cd}=\de_a^c\de_b^d -\de_a^d\de_b^c$ acts like twice the identity on 2-forms. 
The $X$ tensor characterizing the gauge action
satisfies $X_{abcd}=X_{[ab][cd]}=X_{cdab}$. Time-reversal, parity and rotational symmetries 
provide further constraints and make $X$ highly redundant. 
It is convenient to encode its essential information in the diagonal matrix\cite{myers11} $X_{A}^{B}=\diag(X_1,\dots,X_6)$,
where the indices take values in the ordered set $\{tx,ty,tu,xy,xu,yu\}$, so that $X_{tx}^{\si tx}=X_1$, etc.     
Rotational symmetry 
requires $X_1=X_2$ and $X_5=X_6$. Further, in the space of terms that we consider, $X_1=X_5$ and $X_3=X_4$, leaving 
only 2 independent entries out of 6. 
The corresponding matrix for $I_{ab}^{\si cd}$ is the identity, while the one for 
$C_{ab}^{\si cd}$ is $C_A^B=-\tfrac{1}{2}u^3\diag(1,1,-2,-2,1,1)$.   
One can determine the gauge action via matrix multiplication, keeping in mind that any 
time two indices are contracted a factor of 2 appears due to the antisymmetry of the tensors under consideration.
For e.g., $C_{ab}^{\si ef}C_{ef}^{\si cd}$ maps to the matrix $2C_A^EC_E^B=\tfrac{1}{2}u^6\diag(1,1,4,4,1,1)$.  

Given $X$, the relevant equation of motion and corresponding conductivity read\cite{myers11}:
\begin{align}
  &A_y''+\left(\frac{f'}{f}  +\frac{X_5'}{X_5}\right)A_y'+\frac{w^2X_1-q^2fX_4}{f^2X_5}A_y=0 \,,\label{eq:eom_Ay}\\
  &\s(w) = \frac{1}{g_4^2iw}\lim_{q\ra 0}\frac{A_y'}{A_y} \Big|_{u=0}\,, \label{eq:sig} 
\end{align} 
where $A_y(\omega,\b k,u)$ is the Fourier transform in time and $x,y$; $\b k$
is along $x$. We have defined $(\,)'=\pd_u(\,)$, and the rescaled frequency/momentum: 
$(w,q)=\tfrac{3}{4\pi T}(\omega,|\b k|)$. 

\section{Beyond Weyl} 
\begin{figure}
\centering%
\includegraphics[scale=.32]{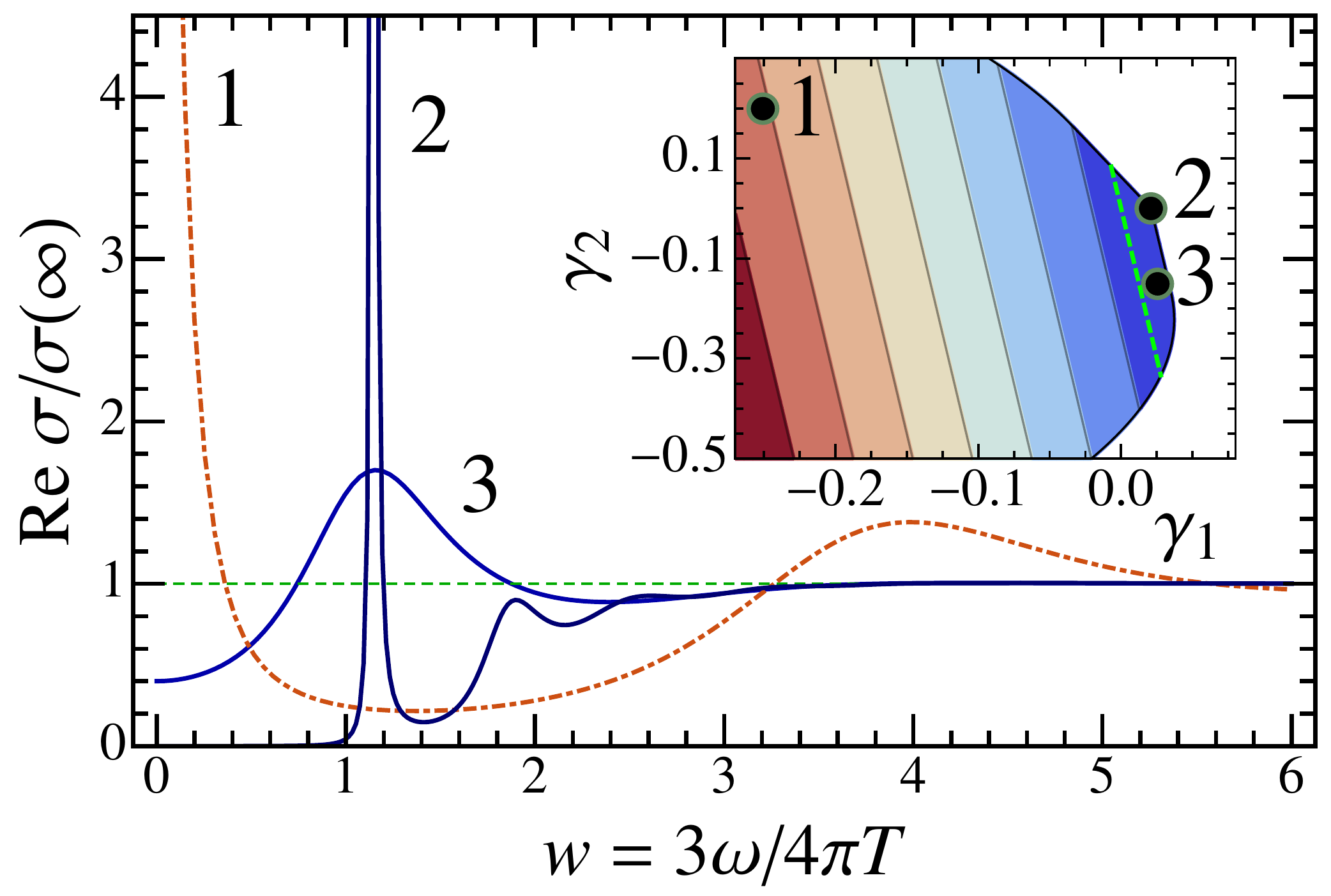}      
\caption{\label{fig:sig_g0}  
Real part of the conductivity for $\g=0$, with the corresponding $(\g_1,\g_2)$ indicated 
in the inset. \underline{Inset}: portion of the allowed parameter space for $\g=0$, and contour plot of  
$\s(0)$, where blue/red indicates small/large values. 
The green dashed lines in the main figure and inset are associated with the self-dual conductivity.   
} 
\end{figure}

In general, we have $X_{i=1,3}=\sum_{n=0}^\infty \Upsilon_i^{(n)}u^{3n}$, where $\Upsilon_i^{(n)}$ is 
a linear combination of couplings appearing at ``level $n$'', i.e.\ from terms involving $n$ Weyl tensors. 
This can be seen from the fact that $C_{ab}^{\si cd}$ scales like $u^3$, so that any term with  
$n$ powers of $C$ makes a contribution $\propto(u^3)^n$ to $X_i$.    
Indeed, from Eq.~(\ref{eq:X}), we easily find $X_1=1+4\g u^3-4(12\g_1+\g_2)u^6+\sum_{n=3}^\infty \Upsilon_1^{(n)}u^{3n}$     
and $X_3=1-8\g u^3-16(3\g_1+\g_2)u^6+\sum_{n=3}^\infty \Upsilon_3^{(n)}u^{3n}$. 
At level $n$, there are $n$ \emph{a priori} independent terms, $\{\tr(C^{n-m})C^m\}_{m=0}^n$. 
(Note that $\tr C=0$.)   
However, since for fixed $n$ we only have  
2 coefficients, $\Upsilon_{i=1,3}^{(n)}$, only 2 terms per level are needed to characterize the charge response.
Focusing on the conductivity,     
Eqs.~(\ref{eq:eom_Ay}) and (\ref{eq:sig}) imply that the latter only depends on $X_1=X_5$; for e.g.\ its d.c.\ value is     
\begin{align} \label{eq:sig_dc} 
  \s(0)g_4^2=\! X_1(1)= 1+4\g-4(12\g_1+\g_2)+\!\sum_{n=3}^\infty \Upsilon_1^{(n)}.      
\end{align}
From the form of $X_1(u)$, we see that
$\s(w)$ will be constant along the lines where $12\g_1+\g_2$ is constant; this is illustrated
for $\s(0)$ in the inset of \rfig{sig_g0}.  

Before proceeding, we need to establish the allowed range for the couplings. 
Following Ref.~\onlinecite{myers11}, we transform the equations of motion of the transverse ($A_y$) and longitudinal ($A_t$)
gauge fields into Schr\"odinger form: $-\pd_z^2\psi_a +V_a\psi_a=w^2\psi_a$, where $dz=du/f$ and   
$a=y,t$ 
labels the transformed $A_{y,t}$ equations, respectively.  
The potentials take the form $V_a=q^2V_{0,a}(u)+V_{1,a}(u)$. As shown in Appendix \ref{ap:bounds},     
it is sufficient to examine the bounds coming from the large-$q$ limit: requiring   
$V_{0,a}(u)\leq 1$ ensures the absence of superluminal modes in the bulk, hence guaranteeing the causality  
of the boundary CFT\cite{myers11}. We find the general form of the potentials to be 
$V_{0,y}=fX_3/X_1$ and $V_{0,t}=fX_1/X_3$. (Their relation   
follows from EM duality, which exchanges the transverse/longitudinal channels.)   
The resulting constraints on the parameters are involved (see Appendix \ref{ap:bounds}).     
One constraint that can be derived simply is $|\g|\leq 1/12$. Interestingly, it was also found 
to hold for the $\g_1=\g_2=0$ holographic  
theory\cite{myers11,hofman09,buchel}, as well as on the CFT side\cite{hofman,myers11}: $\g$ can be  
related to a coefficient of a 3-point function involving 2 currents and 1 stress tensor.   
It remains true in the presence of the infinitely many HD terms considered here.
This follows since the terms that have more than 4 derivatives give 
contributions of the form $u^{3n}, n\geq 2$, to $X_i$, and are thus  
irrelevant for the small $u$ behavior of $V_{0,a}$, which dictates the bound.   

It is instructive to consider the subspace where only $\g_1\neq 0$, in which case $X_1=X_3=1-48\g_1u^6$, such that
the constraints mentioned above are trivially satisfied. An additional constraint results from
requiring $\re\s(w)\geq0$, in particular at zero frequency, which yields    
$\g_1\leq 1/48$. It can be checked that the potentials in this parameter range show no anomalies. 
Further, we have verified that the quasinormal modes, both poles and zeros, of $\s(w)$ remain in the
lower half of the complex frequency plane $\Im w< 0$, confirming the absence of instabilities (Appendix \ref{ap:bounds}). 
This leads 
to the important consequence:
\begin{align}
  \lim_{\g_1\ra-\infty}\frac{\s(0)}{\s(\infty)}=\infty \;\;\;\; \& \;\;\; \lim_{\g_1\ra1/48}\frac{\s(0)}{\s(\infty)}=0\,;
\end{align} 
thus making irrelevant the tight bounds found when including terms up to 4 derivatives\cite{myers11,ritz}, namely   
$2/3\leq\s(w)/\s(\infty)\leq 4/3$. This situation is reminiscent of the fate of the lower bound\cite{kss} on the ratio 
of the shear viscosity to the entropy density, $\eta/s\geq 1/(4\pi)$, which was shown to be violated by HD
terms\cite{brigante07}. Generalizing the calculation of Ref.~\onlinecite{myers11} for the diffusion constant $D$,
we find (see Appendix~\ref{ap:dx})   
that $D$ diverges as $|\g_1|^{5/6}$ as $\g_1\ra-\infty$, while it vanishes linearly (with a log correction) as $\g_1\ra 1/48$.
The corresponding behavior for the charge susceptibility $\chi$ follows from the Einstein relation $\s(0)=\chi D$.  
Thus all the charge response coefficients become unbounded; this is summarized in \rfig{map}b. 

Considering only $\g_{i=1,2}\neq 0$, another remarkable point is that $X_1\equiv 1$ 
on the line $\g_2=-12\g_1$ ($-1/3\leq \g_2\leq 1/12$). In this subspace the conductivity is thus self-dual,   
or frequency independent, $\s(w)=g_4^{-2}$, yet 2 six-derivative terms are present. 
This should be contrasted to the self-dual conductivity found\cite{m2cft} for 
the super-Yang-Mills theory described above, where the holographic gauge action is only $F^2$. In the later
case $X_3\equiv1$, which is no longer true when $\g_i\neq 0$, so that the full current-current correlators 
(at finite momentum) 
will differ along the self-dual line although the conductivity remains invariant. This self-dual subspace, 
which always requires $\g=0$, extends to arbitrarily
HD terms because there are always enough couplings to force $\Upsilon_i^{(n>1)}\equiv0$. This suggests 
that theories with a self-dual conductivity are not as unique as one might infer from the supersymmetric
case\cite{m2cft}. It would be interesting to extend the search on the CFT side\cite{m2cft,lesik}.     

\rfig{sig_g0} shows a range of possibilities for $\s(w)$ beyond the 4-derivative action.    

\section{S-duality and infinities} 
A CFT with U(1) current can be mapped to a new, S-dual, CFT
by gauging the original U(1)\cite{witten}. The corresponding current of the S-dual theory   
is $\hat J_\mu=\epsilon_{\mu\nu\al}f^{\nu\al}/2\pi$, where $f$ is the field strength of the S-dual photon.  
In the case of a complex scalar with particle
current $J_\mu$, this reduces to the usual particle-vortex duality where $\hat J_\mu$ becomes the vortex current.  
In the holographic bulk, S-duality is realized via EM duality\cite{witten,myers11}, and the corresponding S-dual 
conductivity is, as expected, $1/\s$\cite{ws}. 
To obtain the dual action one first introduces a new gauge field     
$\hat A$ by adding a term $\hat A\wedge dF$ to $S_A$, which is innocuous since $dF=0$ by Bianchi.  
The path integral for $F$ can be performed after a shift, leaving behind an S-dual action for $\hat A$:
$S_{\hat A} =\int_x \frac{-1}{8\hat g_4^2} \hat F_{ab} \hat X^{abcd}\hat F_{cd}$, where $\hat F=d\hat A$.
In the subspace studied in this work, 
$\hat X_{ab}^{\si cd}$ is simply 
described by the matrix $\hat X_A^B=(X\inv)_A^B$, i.e.\ the inverse of the original $X$-matrix. 
$X_A^B$ being diagonal leads to $\hat X_i=1/X_i$\cite{myers11}. Now,  
what is the EM dual action when expressed in terms of the metric, 
i.e.\ what is the equivalent of \req{S} for $\hat F$? This can be easily seen using the matrix formalism:
$\hat X=X\inv=[I-8\g C-\cdots]\inv$, where we have omitted the matrix indices. The last expression 
can be expanded as a geometric \emph{series}, $\hat X= \sum_{n=0}^\infty (8\g C+\cdots)^n=I+(8\g C+\cdots)+ \mc O(C^2)$,   
which we can rewrite using covariant tensors (keeping only $\g\neq 0$ to illustrate our point): 
\begin{multline}\label{eq:hatX}
  \hat X_{ab}^{\si cd}=I_{ab}^{\si cd}+8\g C_{ab}^{\si cd}+\tfrac{1}{2}(8\g)^2C_{ab}^{\si ef}C_{ef}^{\si cd} \\
  + \tfrac{1}{2^2}(8\g)^3C_{ab}^{\si ef}C_{ef}^{\si gh}C_{gh}^{\si cd}+ \mc O(C^4)\,. 
\end{multline}
Although benign looking in matrix language, we see that the S-dual action contains an infinite number
of terms. This establishes that \emph{S-duality maps a holographic theory with a finite number
of terms to one with infinitely many terms}. The reverse statement will not hold generally.  
A few remarks are in order.    
First, although the terms in \req{hatX} decay rapidly, 
even when the bound in the original theory is saturated $|\g|=1/12$, what guarantees that the
S-dual theory is valid? A glimpse of the answer was given above: the potentials for the 
transverse/longitudinal modes, $V_{a=y,t}$, are exchanged under S-duality. Thus, a 
valid theory necessarily leads to a valid S-dual.   
Second, in the limit of small couplings, the S-dual theory can be obtained by simply
changing the sign of \emph{all} the couplings, generalizing the 4-derivative result\cite{myers11}. 
Finally, it would be desirable to better understand the connection between the boundary and bulk implementation 
of S-duality for general bulk actions.  
\begin{figure}[!t]  
\centering  
\includegraphics[scale=.31]{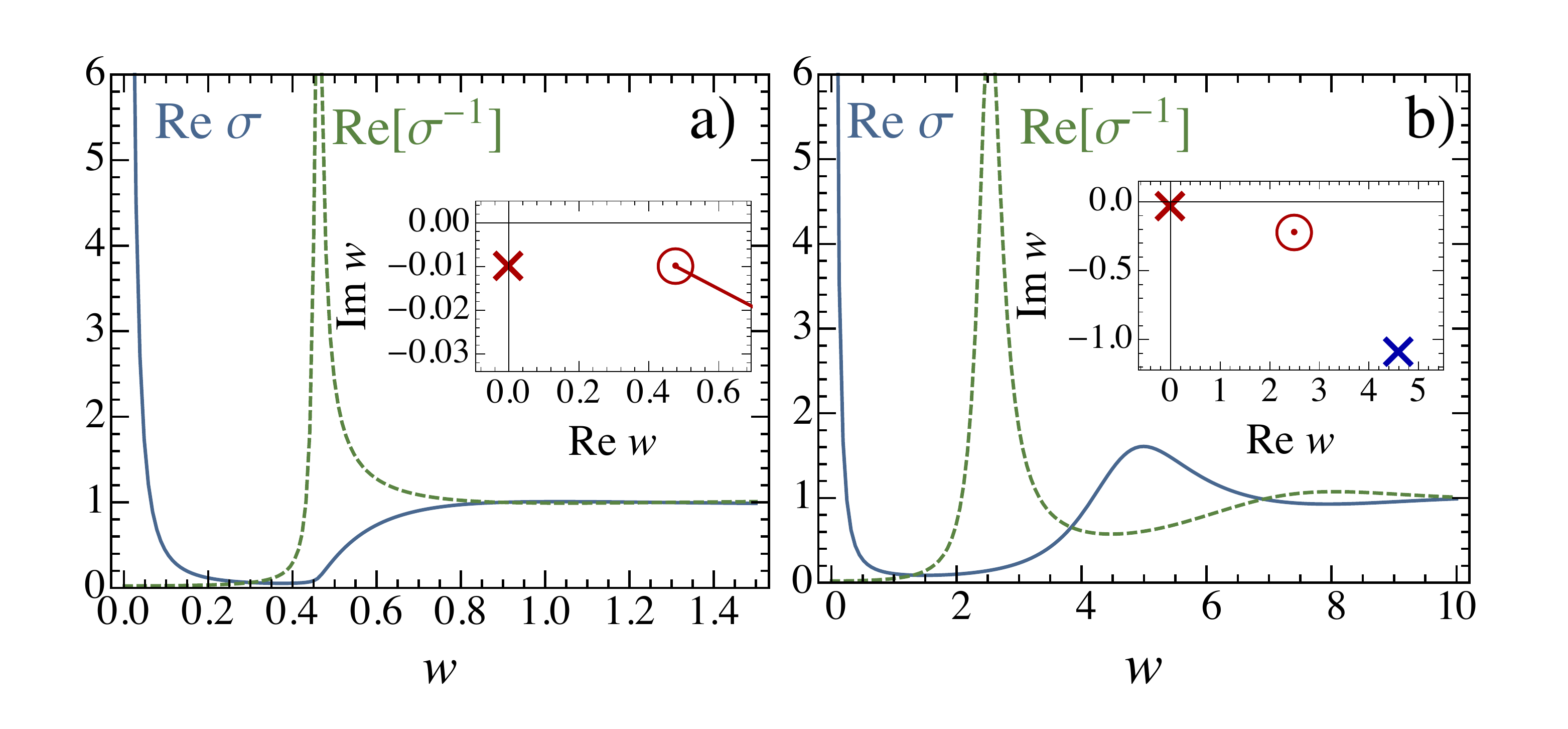} 
\caption{\label{fig:comp} Comparing the conductivity of (a) the quantum critical O($N\ra\infty$) model 
to that of (b) the 6-derivative holographic action with $(\g,\g_1,\g_2)=(0,-1,0)$.  
The inverse conductivities (dashed) also show similarities. \underline{Insets}: analytic structure of $\s(w)$
in complex $w$-plane: crosses/circles represent poles/zeros (the line in (a) is a branch cut). 
} 
\end{figure} 

\section{Connection with O($N$) CFTs} 
\rfig{comp}a shows the conductivity of the O($N$) NL$\s$M in the  
large-$N$ limit at its IR fixed point\cite{damle}. At $N=\infty$, the conductivity 
has a delta-function at zero frequency due the vanishing of interactions. To compare with holography, we introduce
a small broadening factor: $\s_{\rm NL\s M}(\w + i\eta T)$, $\eta\ll 1$. This shifts the pole at $\w=0$  
to $-i\eta T$, matching the result of a calculation including $1/N$ corrections\cite{will-mit,ws}, which found
$\eta\propto 1/N$. \rfig{comp}b shows the holographic result at $(\g,\g_1,\g_2)=(0,-1,0)$.   
In both cases, a sharp Drude-like peak is seen at small $\w$, then
a spectral ``gap'' appears and eventually $\s$ rises and saturates to the $T=0$ value. 
In both cases, the particle-like response\cite{myers11} results from the purely damped pole\cite{ws} below the origin.
Further, the dual responses, $1/\s$, of both the NL$\s$M and holographic results also correlate. 
In that case, $\re[\s\inv]$ 
is suppressed at small frequencies, displaying pseudogap-like behavior; and before saturating shows a  
pronounced peak, which appears because in both cases a zero of $\s$ is 
lurking in the complex $\w$-plane, just below the real axis (\rfig{comp}).        
\begin{figure}
\centering%
\includegraphics[scale=.288]{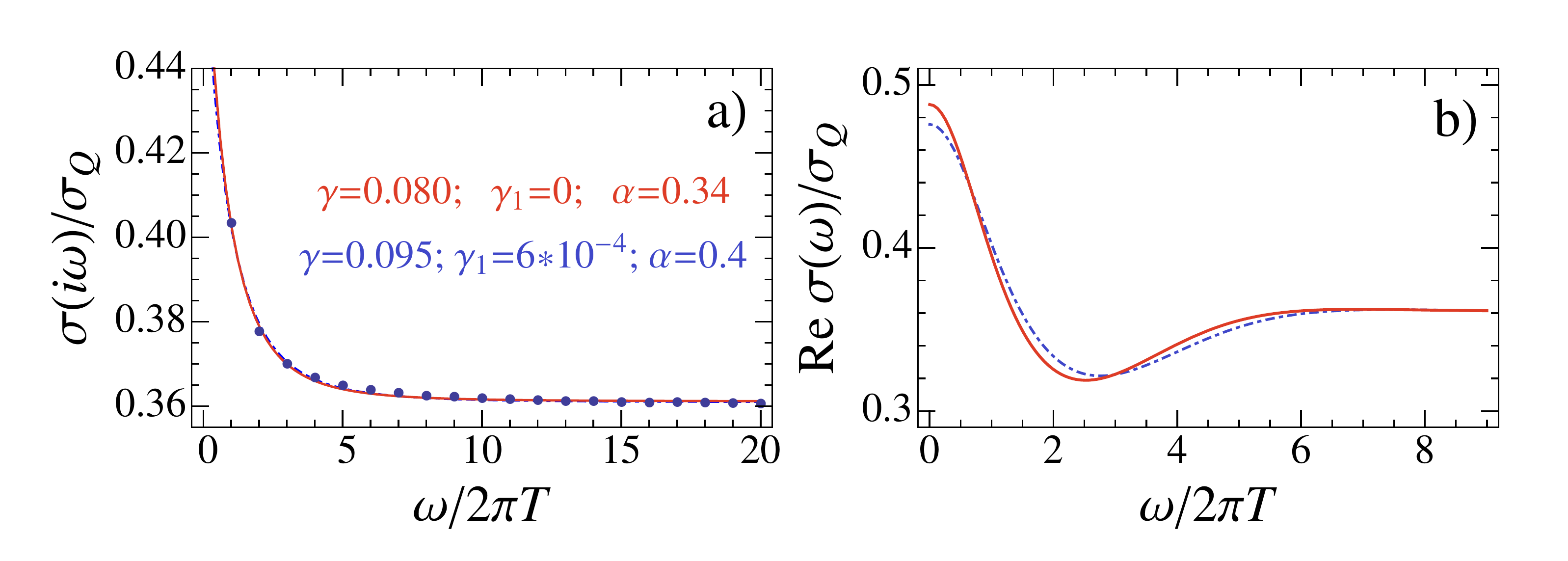}        
\caption{\label{fig:qmc} 
a) The markers correspond to the imaginary-frequency conductivity obtained from large-scale quantum Monte Carlo 
simulations on the O(2) QCP\cite{ads-qmc}. The blue (dot-dashed) and red (solid) curves are the fits to the
data using the Weyl action\cite{ads-qmc} or the one including 6-derivative terms in addition, respectively. 
The values of the fitting parameters are shown.
b) Corresponding real-frequency behavior. 
}  
\end{figure}   
Quantum Monte Carlo simulations have recently provided\cite{smakov,ads-qmc,pollet} high precision estimates of the universal  
charge conductivity of the quantum critical O(2) model in 2+1D, at imaginary frequencies. It was found
that the data can be accurately fitted to $\s$ obtained from the 4-derivative holographic Weyl action, 
with the fitted $\gamma$ obeying the holographic bound\cite{ads-qmc,pollet}. Interestingly, the fit required rescaling 
$\w/T\ra \al\w/T$ on the holographic side, with $\al\approx 0.4$. One might wonder if the rescaling  
can be accounted for by the corrections 
considered above. Focusing on the 6-derivative subspace (and its S-dual), we answer to the negative.
Indeed, the HD terms lead to corrections to $\s$ that are strongest at small frequencies and vanish rapidly 
thereafter. ($\s$ at larger $\w$ can be renormalized when $|\g_1|\gg 1$, but in that case the small $\w$ 
behavior strongly disagrees with the data.) This follows from the fact that the Weyl curvature tensor $C_{ab}^{\si cd}$   
carries relatively more weight in the 
IR (affecting small $\w/T$) compared with the UV, and the effect grows for terms with an increasing number of derivatives.  
The above points to the possible need of quantum gravity renormalizations to account for the rescaling required 
by the O(2) CFT\cite{ads-qmc}. Further, the HD terms do not alter the 
conclusion\cite{ads-qmc,pollet} that $\s$ for the O(2) QCP is particle-like. This is 
summarized in \rfig{qmc}, which shows  
the comparison of the QMC
data with the holographic conductivities obtained from an action involving 6 derivatives and less 
(see also the SM).  

\section{Outlook}    
The previously introduced sum rules for the conductivity\cite{sum-rules,ws,ws2}, 
$\int_0^\infty dw[\re\s(w)-\s(\infty)]=0$, and its S-dual, $1/\s$, hold for the class of
theories considered here, and a general proof is given in Appendix \ref{ap:sr}. The same is true
for the finite momentum relations between the current correlators and those of the S-dual theory\cite{myers11}.  
It would be interesting to   
examine how these, and the other results derived in this work, manifest themselves with other types of HD
terms, such as pure gravity ones. It would also be desirable to  
study the momentum dependence\cite{ws2} of the current correlators and the quasinormal modes\cite{ws,ws2},
as well as extensions to 3+1D and other dynamical exponents\cite{lemos}. 

\indent\emph{Note}:
While preparing the manuscript, we became aware of a related work\cite{bai} that considers a purely-gravitational  
HD term, and finds some conclusions similar to ours.     

\indent \emph{Acknowledgments} -- 
We thank S.~Hartnoll, C.~Herzog, D.~Hofman, R.~Myers, S.~Sachdev and A.~Singh for discussions. Research at Perimeter Institute  
is supported by the Government of Canada through Industry Canada    
and by the Province of Ontario through the Ministry of Research and Innovation.  
  
\onecolumngrid 
\appendix
\section{The infinite family of terms}
We start with the Lagrangian of the gauge field ($L=1$) 
\begin{multline}\label{eq:L_gen}
 \mc L_A g_4^2= -\tfrac{1}{4}F^2+\g C_{abcd}F^{ab}F^{cd} +\g_1 C^2F^2+\g_2 C_{ab}^{\si ef}C_{ef}^{\si cd}F^{ab}F_{cd} \\ 
+\g_{3,1} C^3 F^2 + \g_{3,2} C^2 C_{ab}^{\si cd}F^{ab}F_{cd}
+ \g_{3,3}C_{ab}^{\si ef}C_{ef}^{\si gh}C_{gh}^{\si cd}F^{ab} F_{cd}+ \cdots\,, 
\end{multline}  
where $C^n=C_{ab}^{\si c_1d_1} C_{c_1d_1}^{\si c_2d_2}\cdots C_{c_{n-1}d_{n-1}}^{\si\si\si\si\si ab}$.  
Let us define $\g_{1,1}=\g$ and $\g_{2,i}=\g_i$ ($i=1,2$), relabeling the principal couplings discussed in the main text. 
The above Lagrangian leads to the following $X_i$ functions, which entirely determine the charge
response,
\begin{align}
  X_i(u) &= \sum_{n=0}^\infty X_i^{(n)}(u) \,, \label{eq:Xi_exp}\\
  X_i^{(n)}(u) &=u^{3n}\left(\sum_{m=1}^n \ell_i^{(n,m)}\g_{n,m}\right) =u^{3n}\,\Upsilon_i^{(n)}\,.
\end{align} 
The $\ell_i^{(n,m)}$ are integer coefficients. As was noted in the main text, since at level
$n$ (terms with $n$ curvature tensors) we have $n$ different couplings forming (via linear combinations) only 2 coefficients, 
$\Upsilon_{i=1,2}^{(n)}$, there is a high level of degeneracy. In fact, \emph{we only need 2 terms
per level for $n>1$.} This is because the subspace of couplings for which the 
$\Upsilon_i^{(n)}=\sum_{m=1}^{n}\ell_i^{(n,m)}\g_{n,m}$ stay
constant is $(n-2)$-dimensional ($n$ couplings and 2 linear equations). For example,
one can choose to keep only $\g_{n,1}$ and $\g_{n,2}$ for each $n>1$.  
The conductivity $\s(w)$ only depends on $X_1$, so when examining its behavior it is sufficient to 
restrict oneself to a 1-dimensional subspace per level. In fact, it is simplest to 
keep only $\g_{n,1}C^nF^2$.   

The above analysis has the same purpose as a more formal field redefinition\cite{hofman,myers11} 
analysis, namely the removal of redundancies in the gravitational action. We do not expect that
field redefinitions can further reduce our ``trimmed subspace'' of holographic actions 
because the latter give distinct physical responses. 

\subsection{S-duality and infinite series}
\begin{figure}
\centering
\includegraphics[scale=.45]{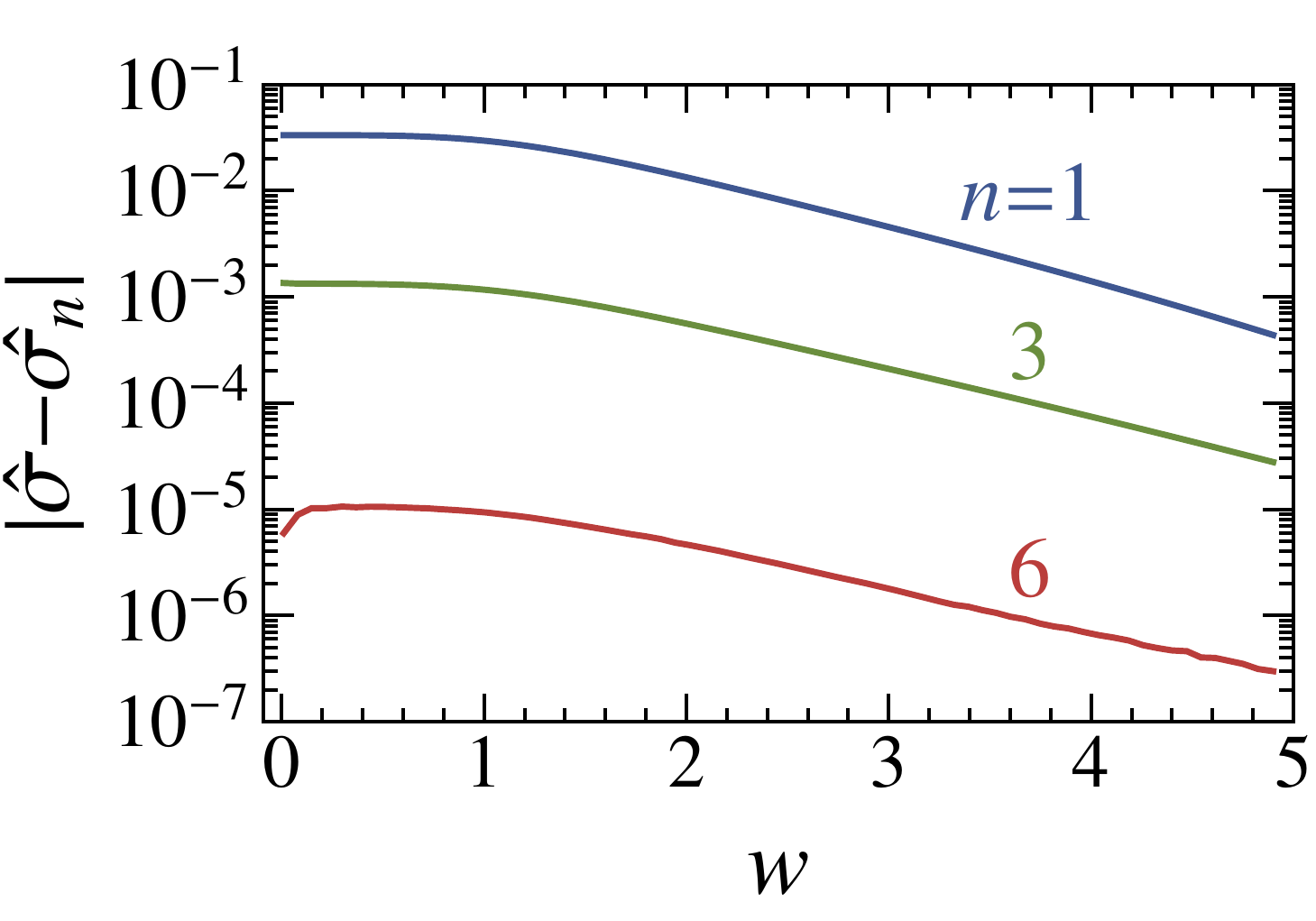}   
\caption{\label{fig:S-trunc} 
Comparison between the exact S-dual conductivity $\hat\s=1/\s$ of the Weyl action at $\g=1/12$, and the one
obtained from truncating the infinite S-dual action to terms with at most $n=1,3,6$ powers of the Weyl tensor, $\hat\s_n$.
$|z|$ denotes the complex norm.   
}  
\end{figure}

It was argued in the main text that S-duality takes an action with a finite number of
terms in the holographic bulk and maps it to one with infinitely many terms. This was illustrated
using the Weyl action, which has only a Maxwell term and $C_{ab}^{\si cd}F^{ab}F_{cd}$. \req{hatX}
gives the X-tensor characterizing the S-dual action, which now contains an infinite string of terms
involving higher and higher powers of $C$. 
We here provide numerical evidence supporting the
analytical argument given in the main text. On one hand we have\cite{ws} $\hat\s=1/\s$, the exact S-dual conductivity;
on the other hand, we obtain the S-dual conductivity by truncating the infinite S-dual action to $n$ powers of the Weyl 
tensor $C$. $\hat\s_1$ results from the S-dual action containing only the Weyl term (in addition to $F^2$),
with its coefficient being of opposite sign to the original theory, and so on. \rfig{S-trunc} shows the   
complex norm $|\hat\s-\hat\s_n|$ for $\g=1/12$, and for $n=1,3,6$. We see that result converges rapidly
to the exact answer as a function of $n$. The agreement is better at larger frequencies, as expected, since
the HD corrections carry little weight in the UV ($u\sim 0$), and thus do not affect much
$\s(w\gg 1)$.

\subsection{Riemann versus Weyl}
We here address the question: Do we get new results by using Riemann instead of Weyl tensors to construct
the holographic action? The answer is no. Let us consider a general HD action made out
of Riemann tensors:
\begin{align}
 \mc L_A= -\tfrac{1}{4}F^2+\al R_{abcd}F^{ab}F^{cd} +\al_1 R^2F^2+ \al_2 R_{ab}^{\si ef}R_{ef}^{\si cd}F^{ab}F_{cd}+ \cdots
\end{align}  
We want to compute its X-tensor. It is again instructive to use the matrix formalism: the matrix associated
with the Riemann tensor is $\mc R_A^B=\diag(\mc R_1^1,\mc R_1^1,\mc R_3^3,\mc R_3^3,\mc R_1^1,\mc R_1^1)$,
with $\mc R_1^1=-(1+u^3/2)$ and $\mc R_3^3=-f=-(1-u^3)$. (Note that we are using a script letter to distinguish
the matrix from the Ricci tensor.) We thus have the matrix form: 
\begin{align}
  X_A^B=I_A^B-8\al_1\mc R_A^B -16\al_1\mc R_C^E\mc R_E^C I_A^B-16\al_2 \mc R_A^C\mc R_C^B\,.
\end{align}
We can readily obtain the corresponding $X_{i=1,3}$ functions, which fully determine the charge response, 
\begin{align}
  X_1 &= (1+8\al-96\al_1-16\al_2)+(4\al-16\al_2)u^3+(-48\al_1-4\al_2)u^6 \,, \\
  X_3 &= (1+8\al-96\al_1-16\al_2)+(-8\al+32\al_2)u^3+(-48\al_1-16\al_2)u^6\,.
\end{align}
We note that $X_1(0),X_3(0)\neq 1$, contrary to the case where only Weyl tensors were used. However, an important
point is that $X_1(0)=X_3(0)$, which is true in general since $\mc R_A^B$ reduces to the identity at $u=0$ (whereas
as the Weyl tensor had the ``nicer'' property of vanishing at $u=0$).   
Note that the region where $X_{i=1,3}(0)=0$ will be forbidden, for otherwise the S-dual action would diverge at the
UV boundary. We can thus factor out $X_1(0)=X_3(0)$ to obtain the new action: 
\begin{align}
  S_A=\int_x-\frac{X_1(0)}{8g_4^2}F^{ab}\t X_{ab}^{\si cd}F_{cd} \,,
\end{align}
where $\t X=X/X_1(0)$. We thus recover an analogous expansion to what we had before, \req{Xi_exp},
\begin{align}
  \t X_i(u) &= \sum_{n=0}^\infty \t X_i^{(n)}(u)\,,\\
  \t X_i^{(n)}&=u^{3n}\, \Phi_i^{(n)}\,,
\end{align}
with $\Phi_i^{(0)}=1$. Therefore, an action containing Riemann tensors can be recast in the same 
form as the action having only Weyl tensors, and will thus not lead to new physics. 
However, an advantage of using Weyl tensors is that the couplings appear in a simpler way
due to the ``separation of scales'' inherent to using the Weyl tensors: a term with $n$ powers
of $C$ will only make contributions scaling like $u^{3n}$, whereas a term with $n$ powers of $R$
will generally contribute to $u^{3n},u^{3(n-1)}$, $u^{3(n-2)}$, etc. 


\section{Bounds on couplings}
\label{ap:bounds}
The bounds on the holographic action can be derived by examining the equations of motion for the gauge field\cite{ritz,myers11}.
Anomalies in those equations lead to either acausal behavior for the boundary CFT or 
instabilities in the holographic bulk, which render the calculation unreliable. We only need to 
consider the equations for $A_t$ and $A_y$, because $A_x'=(q/w)A_t'$ and we have gauge-fixed $A_u=0$.
Our analysis substantially generalizes that of Refs.~\onlinecite{ritz,myers11}, and brings about new insights.  

\begin{figure}
\centering%
\includegraphics[scale=.45]{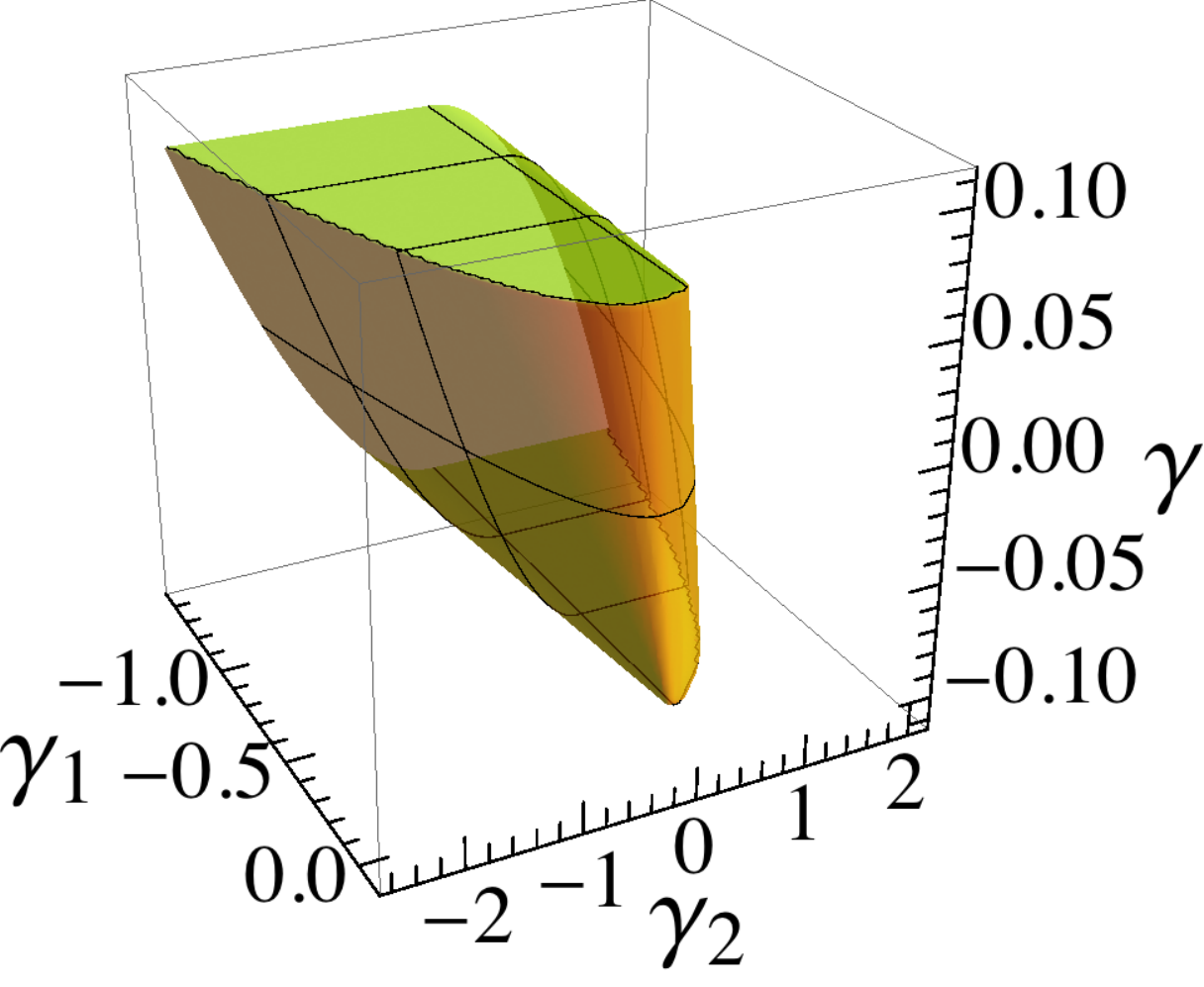}   
\caption{\label{fig:bounds} 
Portion of the allowed parameter space for the $\{\g,\g_1,\g_2\}$ terms, where $\g_{i=1,2}$ are 6-derivative couplings. 
The volume inside the surface is permitted.  
The region is unbounded as $\g_1\ra -\infty$, 
and $|\g|\leq 1/12$.  
}  
\end{figure}

We first bring the 2 equations into a more convenient Schr\"odinger form using the change of variables
$dz/du=1/f$: 
\begin{align}
  -\pd_z^2\psi_a+V_a\psi_a &=w^2\psi_a\,; \\
  A_a=G_a\times\psi_a\, , &\qquad V_a = q^2V_{0,a}+V_{1,a}\,,
\end{align}
where $a=t,y$ (it is not summed over), 
$V_a$ are the potentials, and $G_a$ are auxiliary functions used to remove terms linear in $\pd_z$:
$G_t=(X_1/X_3^2)^{1/2}$ and $G_y=1/X_1^{1/2}$. The potentials have been decomposed into a part that depends on
momentum, $q^2V_{0,a}$, and one that does not, $V_{1,a}$. 
The key constraints in the analysis come from the limit of large momentum, so that $V_a\approx q^2V_{0,a}$; they
read\cite{myers11} (see below for further details)  
\begin{align}\label{eq:bounds_V0}
  V_{0,a}(u) \leq 1\,,\quad {\rm for}\;\; 0\leq u \leq 1\,.
\end{align}
We have found that in general
\begin{align}
  V_{0,t}=fX_1/X_3\,, \qquad V_{0,y}=fX_3/X_1\,. 
\end{align}
The two potentials are simply related to each other: $V_{0,y}=V_{0,t}\big|_{X_i\ra\hat X_i}$, and vice-versa. 
This must be true because S-duality exchanges the longitudinal/transverse channels ($t\leftrightarrow y$),
and maps $X_i$ to $\hat X_i=1/X_i$. 
Now, if the potentials were to exceed unity, this would lead to the appearance of superluminal modes, and a corresponding
violation of causality in the boundary theory.
Let us first review the situation when $\g$ is finite, and all HD terms vanish. 
In that case the constraint is\cite{myers11} simply $|\g|\leq 1/12$. It turns it can be derived from the near-boundary, 
$u\sim 0$, behavior alone. Indeed, a Taylor expansion yields: $V_{0,a}=1-(1\mp 12\g)u^3+\mc O(u^6)$, 
for $a=t,y$, respectively.   
Interestingly, when in addition $\g_1,\g_2$ are finite, the constraints from \req{bounds_V0} can also be derived analytically.
However, due to their complexity, we find it more advisable to plot the allowed region, which is shown in \rfig{bounds}.    
One constraint that survives is $|\g|\leq1/12$. This is the case for the entire family of HD terms 
considered in this work. 
As mentioned in the main text, such a conclusion comes about because the contributions
from the higher derivative terms become negligible at small $u$ compared to that of the $\g$ term. 
We note that the $|\g|<1/12$ constraint is not tight, in the sense that for some some values of $(\g_1,\g_2)$, 
the allowed $\g$-range is strictly contained in $[-1/12,1/12]$. This can indeed be seen in \rfig{bounds}: the allowed 
volume is \emph{not} a (generalized) cylinder in the $\g$-direction.  

Just like the $\g\neq0$ subspace, the allowed region in \rfig{bounds} is topologically connected.  
However, an important distinction exists: it is \emph{unbounded}. This was already pointed out 
in the main text, where it was noted that in the special case when $\g_1$ is the only non-zero coupling, $X_1=X_3=1$ for all $u$,
trivially satisfying the constraints \req{bounds_V0} for all $\g_1$. Requiring further that the 
real part of the conductivity be positive enforces $\g_1<1/48$. In general when $\g,\g_{i=1,2}\neq 0$,
the $\g_1=\text{constant}$ cross-section of the allowed volume \rfig{bounds} grows as $\g_1\ra-\infty$.
Taken in conjunction with the analysis below,
which does not rule out this region, this implies the important conclusion that the 
d.c.\ conductivity becomes unbounded as $\g_1\ra -\infty$. Indeed, recall that 
$\s(0)g_4^2= 1+4\g-4(12\g_1+\g_2)+\sum_{n=3}^\infty \Upsilon_1^{(n)}$.
Via S-duality, this yields a conductivity ($\hat\s=1/\s$) that goes to zero in the d.c.\ limit as $\g_1\ra-\infty$.

The constraints \req{bounds_V0} also imply $V_{0,a}\geq 0$. Indeed, we have  
$V_{0,t}V_{0,y}=f^2=(1-u^3)^2$, implying that $V_{0,a=t,y}$ cannot vanish for $0<u<1$, since they are 
both bounded from above (by virtue of \req{bounds_V0}) and $V_{0,a}(0)=1$. The latter condition can
be seen from the general form, $X_i=\sum_{n=0}^\infty \Upsilon_i^{(n)}u^{3n}$, with $\Upsilon_i^{(0)}=1$.  
(Naturally, the potentials we consider are continuous.)   

\subsection{Small momentum}
In the limit of small $q$, the potentials $V_{1,a}$ dominate, and one must ensure
that no anomalies arise from them. Indeed, these can develop minima that can potentially
lead to ``bound'' states, and corresponding unstable QNMs (in the upper complex half-plane).
Such an instability of the neutral holographic plasma would render the calculation of the charge
response unreliable\cite{myers11}. We will now show that when $V_{0,a}\leq 1$ this will not occur, by 1) looking 
for bound states within a WKB approximation, 2) directly examining the QNM spectrum.

We derive the $V_{1,a}$ potentials in general for the theories considered in this work: 
\begin{align}
  V_{1,t}(u) &= \frac{f}{4X_1^2}\left[3f(X_1')^2 -2X_1(fX_1')'\right]\,; \label{eq:V1t} \\ 
  V_{1,y}(u) &= \frac{f}{4X_1^2}\left[-f(X_1')^2 +2X_1(fX_1')'\right]\,, \label{eq:V1y} \\
            &= V_{1,t}\Big|_{X_1\ra \hat X_1=X_1\inv}\,.  \label{eq:V1_Sdual} 
\end{align}
It might seem surprising at first that the potentials have the same form and only depend on
$X_1$ and not $X_3$. The facts follow from S-duality, as shown for the $q$-dependent 
part of the potential in \req{V1_Sdual}.    
Indeed, the duality interchanges the transverse/longitudinal channels, $y\leftrightarrow t$,  
and at the same time maps $X_1$ to $\hat X_1=1/X_1$. Combining these results with those for $V_{0,a}$
we conclude the simple relation for the full potentials: 
\begin{align}
  V_t=V_y\big|_{X_i\ra \hat X_i}\,, 
\end{align} 
and vice-versa. 

\begin{figure}   
\centering
\subfigure[]{\label{fig:V1t_gam0} \includegraphics[scale=.38]{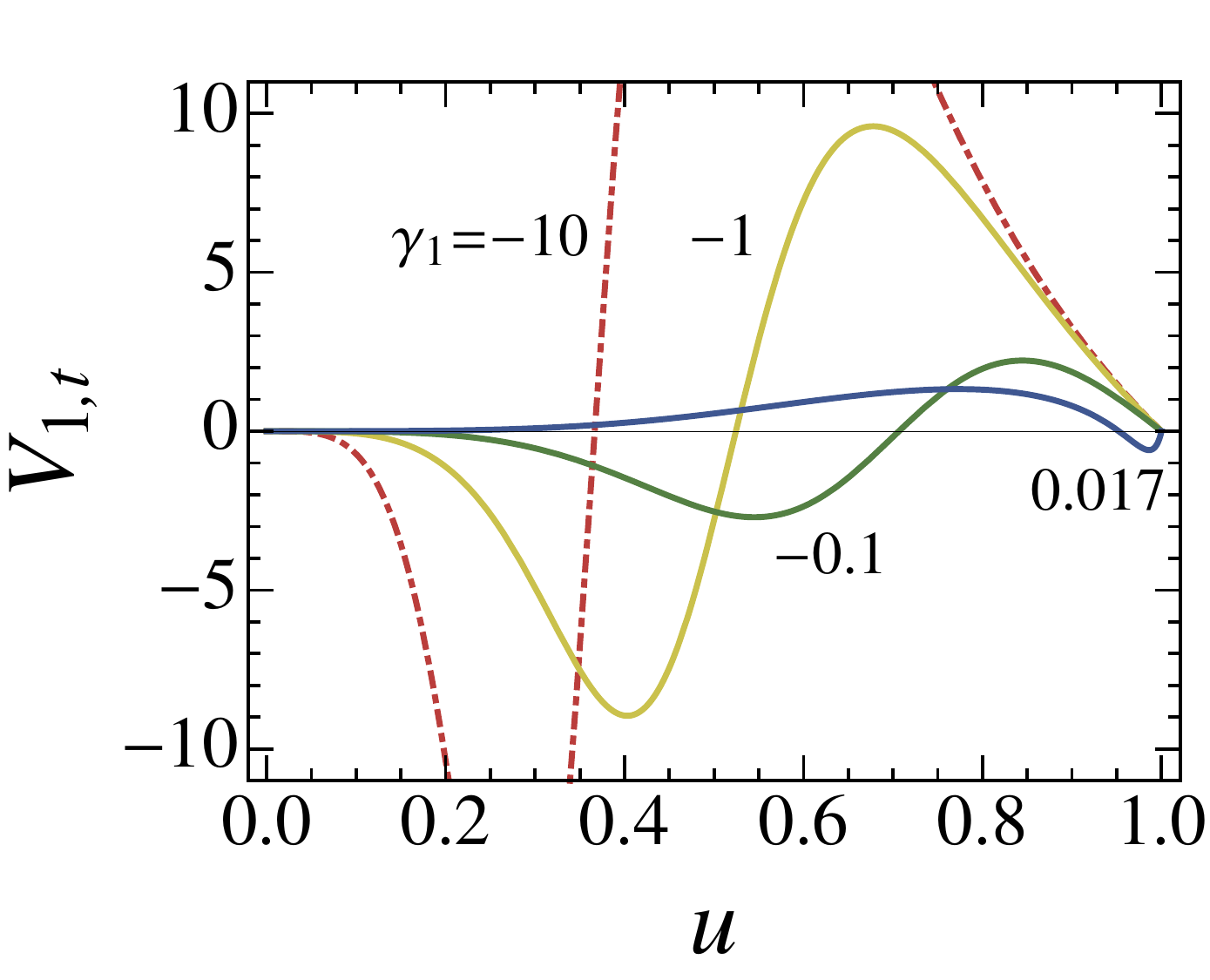}}       
\subfigure[]{\label{fig:V1y_gam0} \includegraphics[scale=.38]{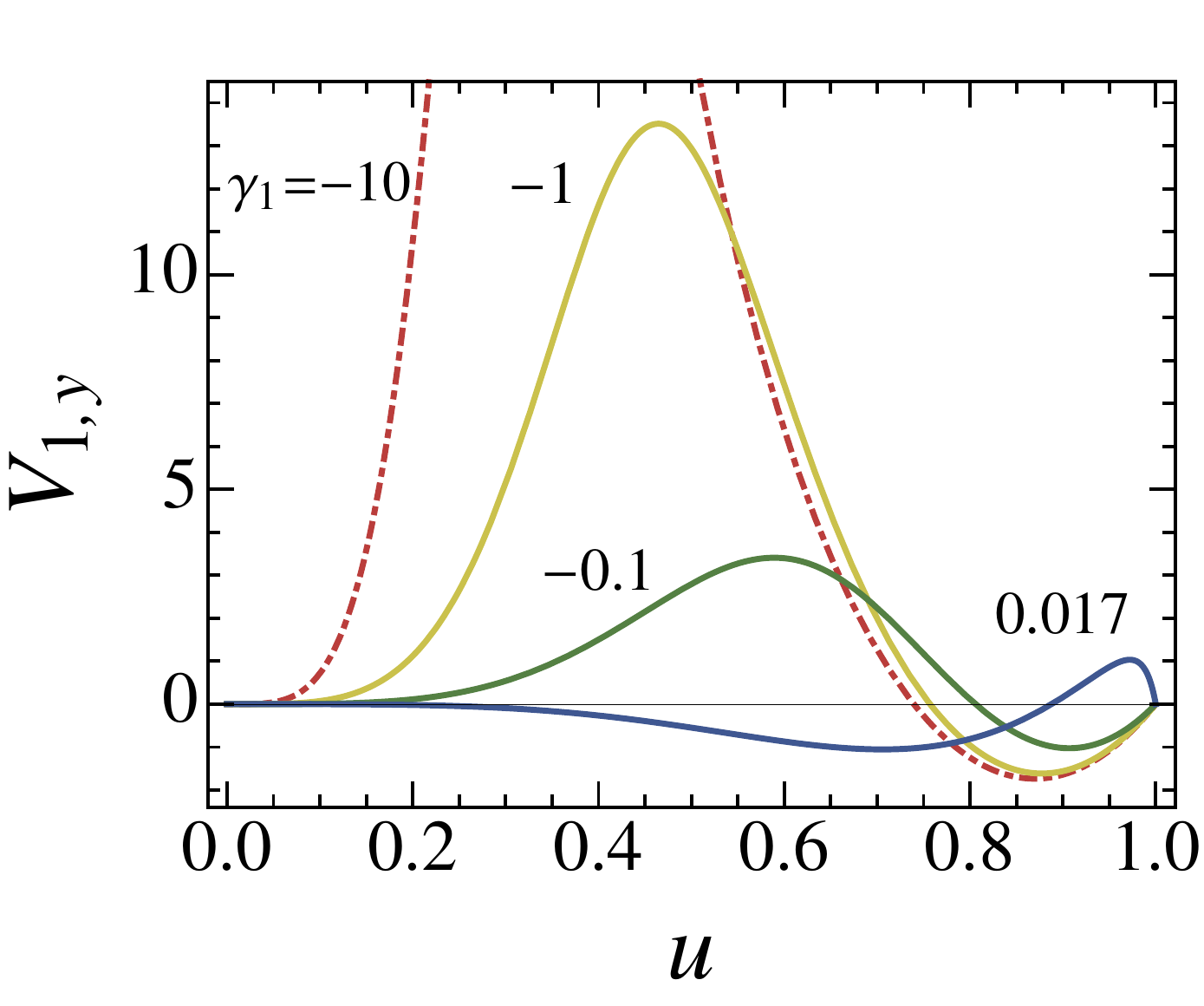}} \\       
\subfigure[]{\label{fig:nt_gam0} \includegraphics[scale=.38]{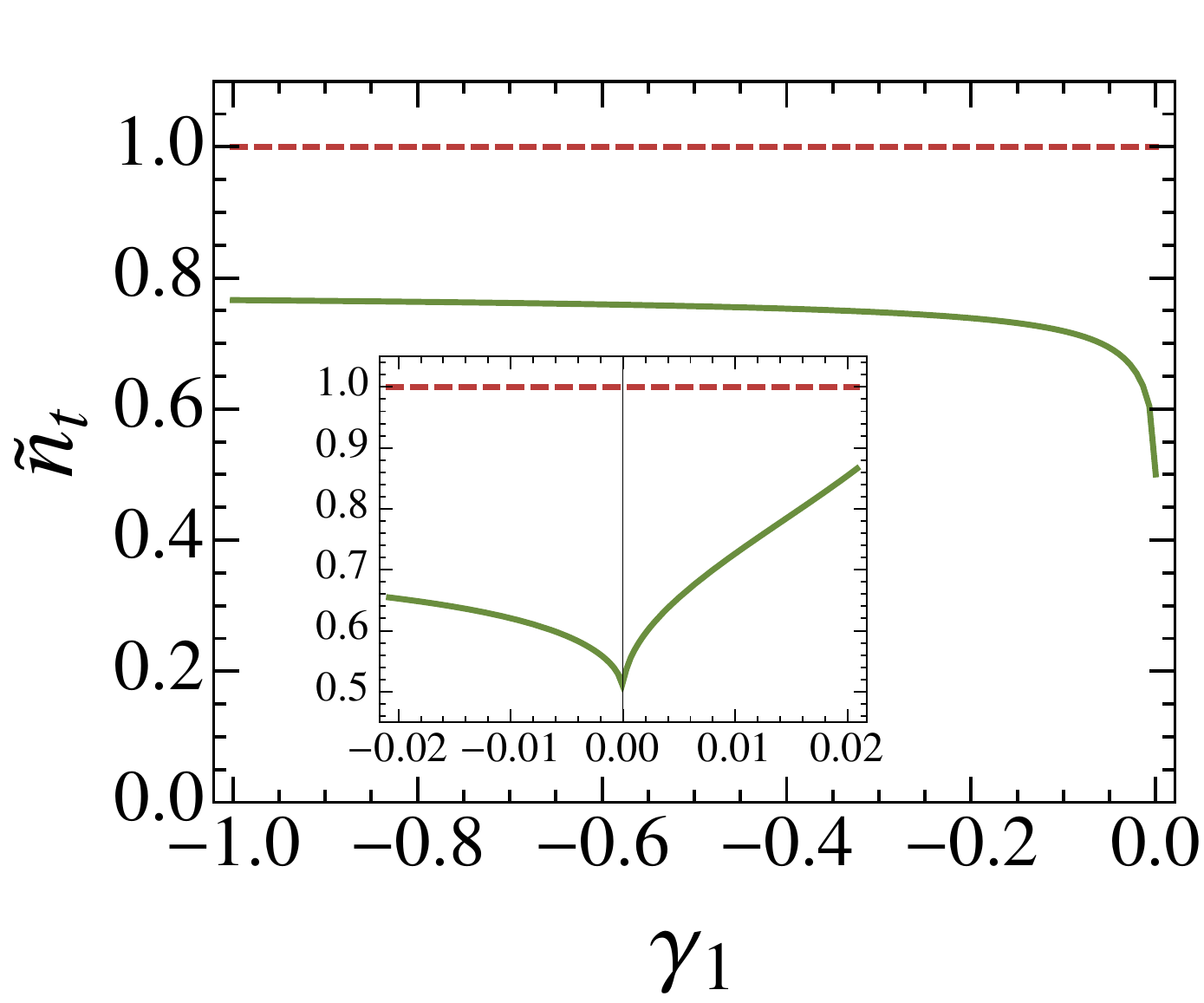}} 
\subfigure[]{\label{fig:ny_gam0} \includegraphics[scale=.38]{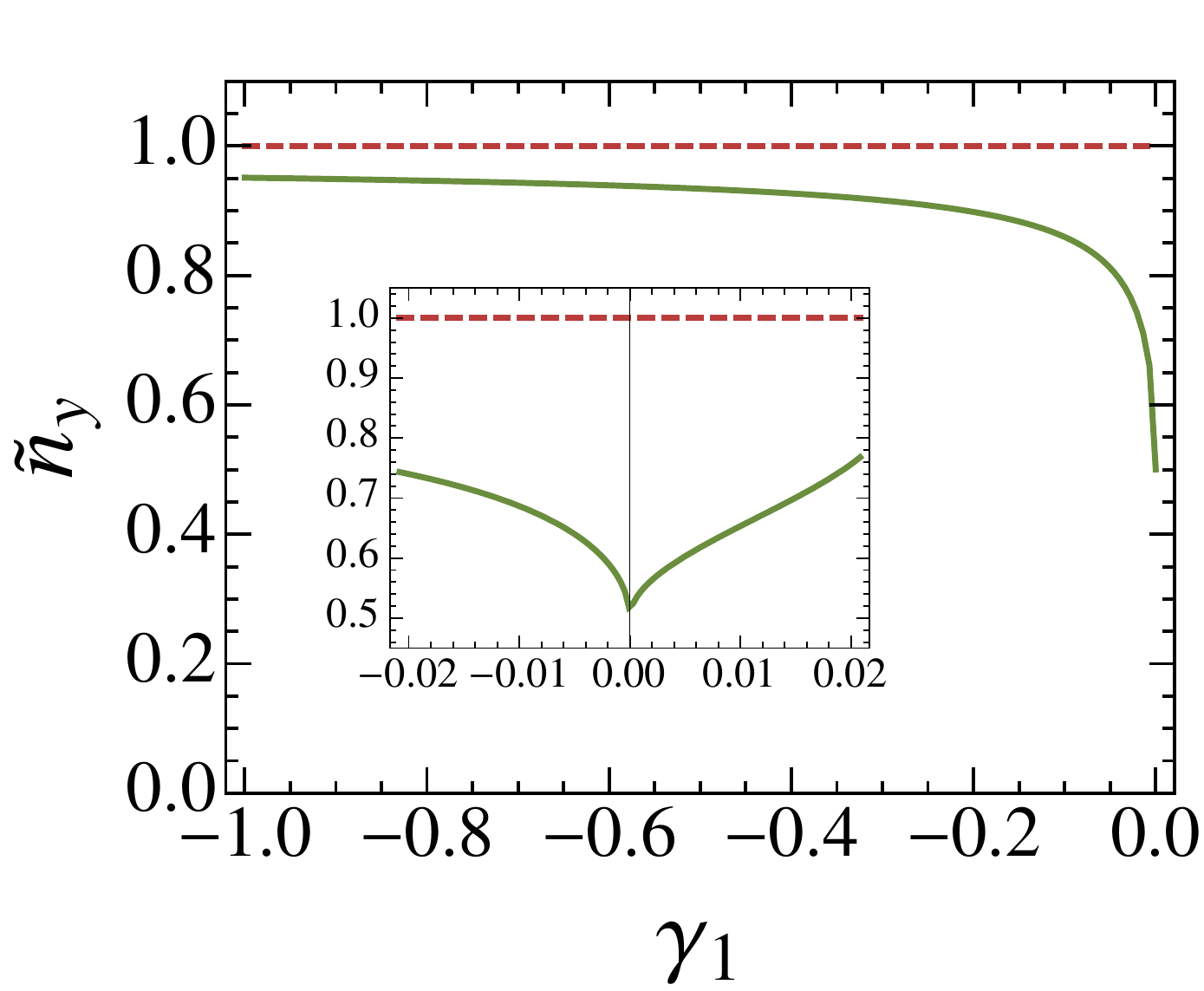}}  
\caption{\label{fig:V1} Panels (a) and (b) show the potentials $V_{1,a=t,y}$
for $\g=0=\g_2$ and different values of $\g_1$. As shown in (c) and (d), the
negative potential wells are never sufficiently deep to develop bound states
when the couplings obey the bounds coming from \req{bounds_V0}. Bound states, and hence
instabilities, would appear if
$\t n_a$ were to reach 1, indicated by the red dashed lines.   
} 
\end{figure}

A bound state can occur when a sufficiently deep minimum develops in the potential 
in a region where $V_{1,a}<0$. Given a region defining a negative potential well, we  
want to determine whether it can capture a bound state. One way to proceed is to use
the WKB approximation often used in quantum mechanics. It has indeed been empirically found by previous
works that this is sufficient\cite{star1,myers11}, and we will see that it remains so in our case. 
The WKB condition for a bound state to appear is:
\begin{align}\label{eq:V1_wkb}
  (n-1/2)\pi = \int_{u_1}^{u_2}\frac{du}{f}\sqrt{-V_{1,a}}\,,
\end{align}
where $n$ is a positive integer $n\geq 1$.
We have changed variables back from $z$ to $u$, and the interval $u_1< u < u_2$ defines the negative well.  
Defining $I_a$ to be the integral on the r.h.s.\ of \req{V1_wkb}, we introduce the quantity 
$\t n_a=\pi\inv I_a+1/2$. We see that the condition to get a bound state becomes $\t n_a\geq 1$. 
As an example, let us examine the case $\g=0$. Since the potential $V_{1,a}$ only depends on $X_1$, it is sufficient to
keep only $\g_1$ and set $\g_2=0$ in order to probe the entire set of possibilities. We plot
$V_{1,t},V_{1,y}$ in \rfig{V1t_gam0} and \rfig{V1y_gam0}, respectively. We observe that the negative potential
well can be bounded on either side by the horizon, or the UV boundary $u=0$. In \rfig{nt_gam0} and \rfig{ny_gam0}  
we show the calculated $\t n_a$ for a subset of the range of $\g_1$ values allowed by the large momentum 
constraints, \req{bounds_V0}, 
i.e.\ $\g_1<1/48$. We see that $\t n_a$ never reaches 1. In fact, we find $\lim_{\g_1\ra -\infty} \t n_t\approx 0.7887$,
$\lim_{\g_1\ra 1/48} \t n_t\approx 0.8660$, and for $\t n_y$: $\lim_{\g_1\ra -\infty} \t n_y\approx 0.9745$,
$\lim_{\g_1\ra 1/48} \t n_y\approx 0.7681$. All these values are less than unity,
showing that no bound states can form, and that
no additional constraints come about from considerations of $V_{1,a}$. The same conclusion remains true when $\g\neq 0$, 
although we do not show the details of the analysis. We note that the asymptotic values of $\t n_a$ 
as $\g_1\ra-\infty$ do not change in the presence of $\g$, as expected since in that limit the $\g_1$ term dominates 
everywhere except in the $u\ra 0$ limit.  

We note that bound states cannot appear when a finite momentum $q$ is turned on because
$V_{0,a}\geq 0$ in the region where \req{bounds_V0} holds, as proved above. In other words,
a finite momentum can only make the negative potential wells less deep.  

\subsection{Quasinormal modes}


\begin{figure}   
\centering
\subfigure[]{\label{fig:d-qnm_gam0} \includegraphics[scale=.42]{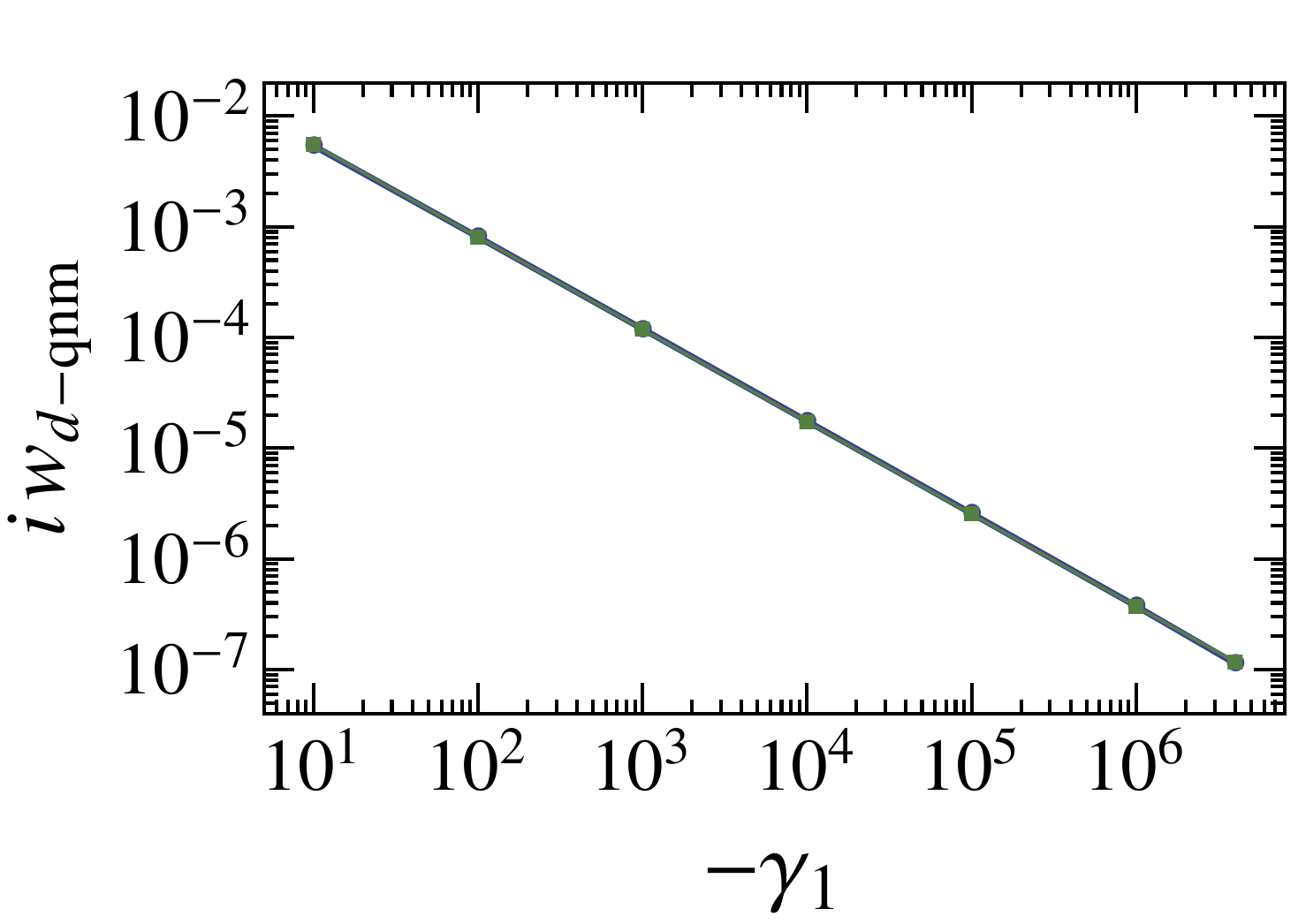}}       
\subfigure[]{\label{fig:zero-qnm_gam0} \includegraphics[scale=.345]{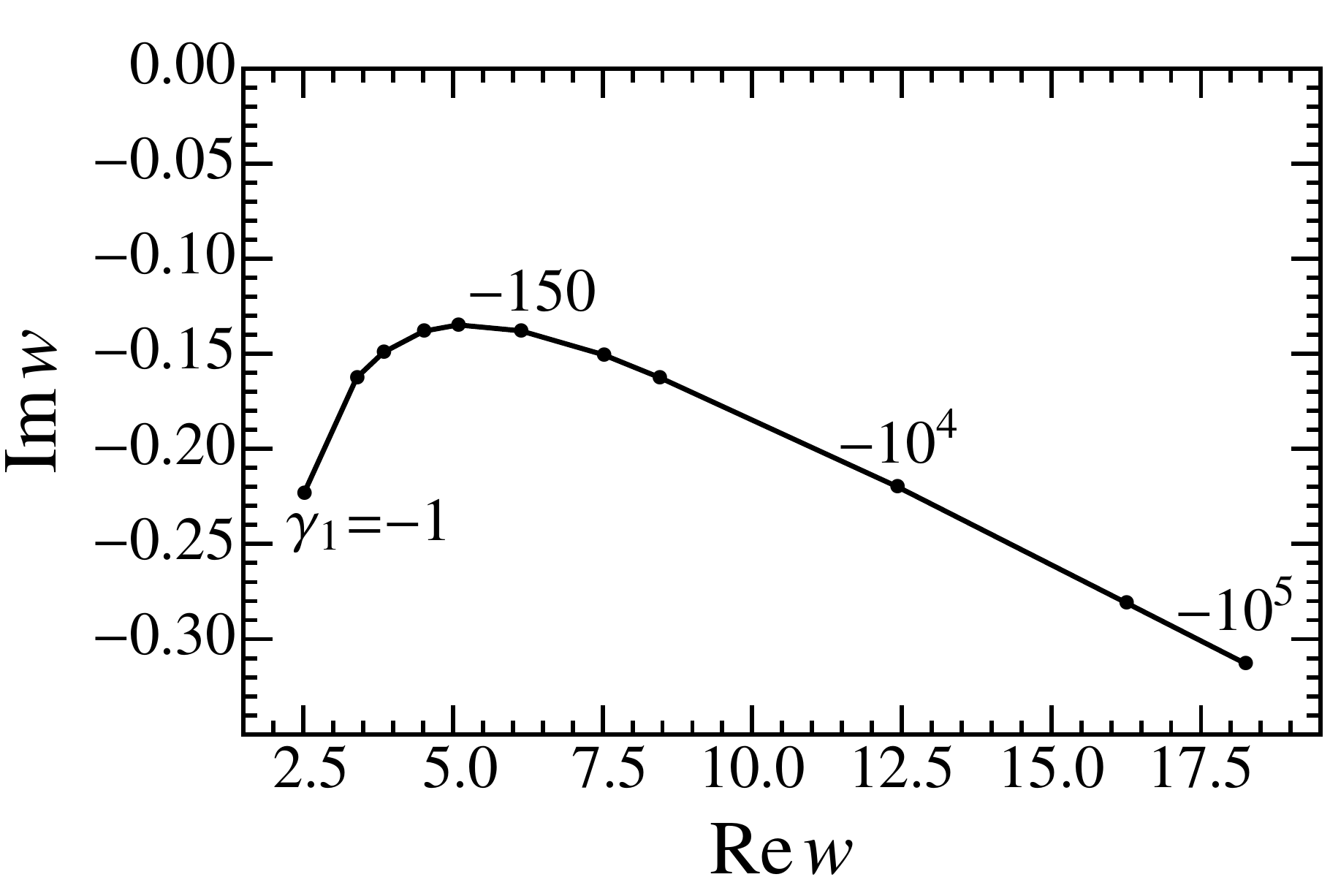}}
\caption{\label{fig:qnm_gam0} a) Location of the purely damped QNM for $\g_1\neq0$, $\g=\g_2=0$. Superimposed on the
numerical data (blue circles) is a power-law fit (green squares) of the form $a/|\g_1|^{0.8345}$.  
b) Location of the next QNM closest to
the real axis, a zero, shown in the complex $w$-plane. It is plotted for $-\g_1=1,5,10,25,50,150,500,10^3,10^4,5\times 10^4,10^5$.
}  
\end{figure}

To confirm the above analysis of the bounds, we examine the quasinormal mode (QNM) spectrum\cite{star1,ws}.   
We indeed find that as long as one stays confined to the subspace of parameters defined by
\req{bounds_V0}, all the QNMs stay confined to the lower half-plane, $\im w\leq 0$.
For example, let us take the 6-derivative action studied in the previous subsection, $\g=\g_2=0$.
As $\g_1\ra-\infty$, the peak of $\re\s$ at small frequencies becomes sharper, continuously approaching a 
delta-function. This results from a pole QNM approaching the origin. Such a  
QNM was labeled D-QNM in Ref.~\onlinecite{ws2}, because of its purely damped nature and its formal relation to the
Drude conductivity. It takes the form $w_{\rm d-qnm}=-i\la$, with $\la>0$. We plot $iw_{\rm d-qnm}$  
in \rfig{d-qnm_gam0}, which shows that the pole indeed asymptotically approaches the origin as $\g_1\ra-\infty$. 
The numerical data strongly suggests a power law $w_{\rm d-qnm}=-i a/|\g_1|^{0.8345}$, where $a\approx 0.0377$,
at least over the 5 decades we have studied: the fit is excellent as shown in \rfig{d-qnm_gam0}.  
Another potentially ``dangerous'' QNM is the zero near the real axis, which is shown in \rfig{comp}b. 
Its path in the complex $w$-plane is shown in \rfig{zero-qnm_gam0}: we see that it never threatens to
reach the real $w$-axis. 

\section{Relating the real and imaginary time responses} 
\label{ap:imag}
We discuss some qualitative features of the conductivity on the real- and imaginary-frequency axes,
the latter being particularly important when analyzing quantum Monte Carlo results.  
In particular, we are interested in understanding how the particle- and vortex-like
features on the real axis correlate with the imaginary-frequency conductivity in the presence of HD terms.
At the level of the Weyl action, it was found\cite{myers11} that when $\g>0$, $\s$ has a particle-like 
peak at small and real frequencies,
whereas for $\g<0$ a vortex-like dip results. On the imaginary-frequency axis these correspond to 
convex/concave behavior, respectively. Recent quantum Monte Carlo simulations\cite{ads-qmc,pollet} on the O(2) QCP 
have found the former to occur. A direct comparison with holography was then used\cite{ads-qmc,pollet} to infer that the 
charge response of the QCP is of particle-like type. We will argue here that such conclusions are robust to the inclusion
of HD terms. 

We start with the following observation, relating $\s$ at imaginary frequencies $i\varpi$ 
to real ones $w$; it characterizes the response near $w=0$:
\begin{align}
  \pd_\varpi^2 \s(i\varpi)\big|_{\varpi=0} = 
  \begin{cases}
    > 0: & \text{ particle-like, i.e.\ }   \pd_w^2\s(w)\big|_{w=0} < 0 \\
    < 0: & \text{ vortex-like,\phantom{il} i.e.\ }  \pd_w^2\s(w)\big|_{w=0} > 0 
  \end{cases}
\end{align}
where $\varpi$ is real, so that $i\varpi$ is along the imaginary axis. The real-frequency conditions on the r.h.s.\ 
state whether there is a local maximum/minimum at $w=0$. ($w$ is taken to be real, unless otherwise specified.)
These conditions simply follow from analytic continuation, $w\ra i\varpi$, and thus remain valid irrespective
of the choice of holographic action. At the level of the action containing only the $\g$ term,
the possibilities on the \emph{entire} imaginary-frequency axis are quite limited: for all $\varpi\geq0$,   
$\sgn[\pd_\varpi^2\s(i\varpi)]=\sgn\g = -\sgn[\pd_\varpi\s(i\varpi)]$, so that the sign of the first and second
derivatives is constant. For e.g.\ when $\g>0$, the imaginary-frequency conductivity is monotonously decreasing
and convex for all $\varpi\geq0$.
This uniform behavior does not always occur when HD terms are included, although it is by far the dominant
one, as we now argue.  

\begin{figure}
\centering%
\includegraphics[scale=.48]{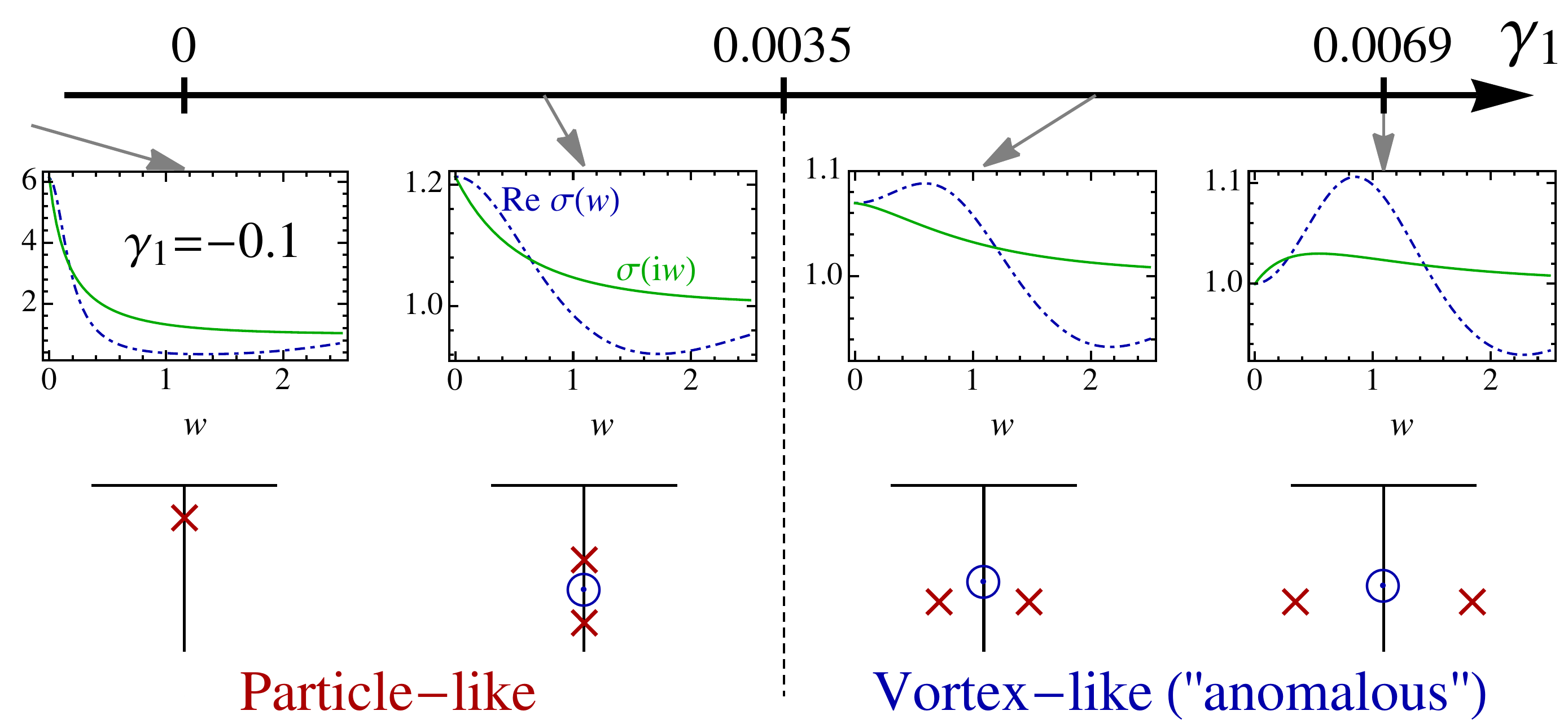}       
\caption{\label{fig:im-frequ}  
Behavior of the conductivity for real (blue, dot-dashed) and imaginary (green, solid) frequencies as a 
function of $\g_1$, with $\g=1/12$ fixed. From left to right: $\g_1=-0.1, 0.0025, 0.0055, \text{ and } 1/144\approx 0.0069$.
(We have the constraint $\g_1<1/144$.) 
The corresponding QNM spectra are shown in the lower part, with the crosses/circles representing poles/zeros. 
A transition between particle- and vortex-like responses is observed at $\g_1\approx 0.0035$. 
The ``anomalous'' vortex-like response arises because satellite poles occur near the dominant zero QNM.
Such a situation did not arise in theories with 4 derivatives and less. 
}   
\end{figure}

To illustrate our point, let us consider terms up to 6 derivatives in \req{L_gen}, with $\g=1/12$ fixed. As was mentioned above,
we only need to tune $\g_1$, keeping $\g_2=0$, in order to examine the possible behavior for the conductivity.
When $\g=1/12$, the bound on $\g_1$ is $\g_1<1/144=\g_{1,\rm max}$ (which is obtained by requiring $X_3(1)\geq0$). 
\rfig{im-frequ} shows representative behavior for the conductivity, along both the real (blue and dot-dashed) 
and imaginary (green and solid) frequency axes. For most of the allowed range, $\s$ is particle-like,
and there are no surprises on the imaginary axis: $\s(i\varpi)$ is monotonously decreasing and convex 
for all imaginary frequencies $\varpi\geq0$. 
On the other hand when, when $\g_1>\g_{1,c}\approx 0.00348$ (note that $\g_{1,c}\approx \g_{1,\rm max}/2$), the 
small frequency conductivity becomes vortex-like at small frequencies. In \rfig{im-frequ}, we have specified that this 
vortex-like behavior is of an ``anomalous'' type. Indeed, the dip at small frequencies rapidly gives 
way to a larger particle-like feature. Associated with such a crossover, the imaginary-frequency conductivity 
goes from being concave at small frequencies to convex at larger frequencies. The inflection points can indeed
be seen in the right part of \rfig{im-frequ}. We further note that for $\g_1$ near the bound, the imaginary-frequency 
conductivity is even \emph{non-monotonous}, as shown in the rightmost panel of \rfig{im-frequ}. 
This must be the case because when   
$\g_1=\g_{1,\rm max}=1/144\approx 0.0069$, $\s(i\varpi=0)=\s(i\varpi=i\infty)$, 
yet the conductivity is not self-dual since $\g\neq0$, hence it cannot be constant 
in the complex plane. (We note that $\s(0)=\s(i0)$ and $\s(\infty)=\s(i\infty)$.) 

This anomalous behavior, which takes place over a very narrow range of the allowed 
values of $\g_1$, occurs because 
of the tension between the 4-derivative term parameterized by $\g$ and the 6-derivative one. This can be seen from the
d.c.\ conductivity, $\s(0)g^2_4=1+4\g -48\g_1$: there exists a tension when $\g$ and $\g_1$ are of the same sign,
because in that case one term strives to make the conductivity vortex-like, whereas the other strives for a particle-like 
response. When $\g=1/12$ as in our example, $\g_1$ only succeeds in making a very weak vortex-like $\s$ when it
is positive; however, when the tension is relieved for $\g_1<0$, the conductivity is particle-like, from the
``combined effort'' of both terms. 

As is often the case, extra insight can be gained when examining the QNM spectrum of $\s$, which is 
shown in \rfig{im-frequ}. The particle/vortex-like conductivities are characterized\cite{ws} by
a purely imaginary pole/zero QNM closest to the real axis (which was called the D-QNM in Ref.~\onlinecite{ws2}).  
We see it is indeed the case in \rfig{im-frequ}.
What makes the vortex-like response ``anomalous'' is the presence of the satellite poles whose imaginary part is
nearly equal to that of the zero. This reduces the domain of influence of the zero, the D-QNM.  
This is especially true for $\g_1\approx \g_{1,c}$, because the poles are very near the imaginary axis, 
in preparation for the spectral transition at $\g_{1,c}$ where the poles attach to the imaginary axis and
the D-QNM becomes a pole. 

\subsection{Comparison with quantum Monte Carlo}
The holographic fit to the imaginary-frequency quantum Monte Carlo data of Ref.~\onlinecite{ads-qmc} is
shown in \rfig{qmc}. It was found that $\g=0.08$, together with a rescaling of the frequency axis, $\w/T\ra \al\w/T$
with $\al=0.35$, gave an excellent fit to the data. We have repeated the fitting procedure to the data with the inclusion
of the $\g_1$ 6-derivative term. The resulting fit is shown in red in \rfig{qmc}. We see that the imaginary-frequency
dependence is essentially the same as that of the previous fit, and the corresponding real-frequency behavior 
changes very little from the inclusion of $\g_1$. We have repeated the fit with many different initial guesses
for the value of the fitting parameters, and the shown fit has the small estimated variance\footnote{The estimate variance is
defined as $(1/\t N)\sum_{n=1}^N[\s_{\rm fit}(iw)-\s_{\rm qmc}(iw)]^2$, 
where $\t N =\text{(\# data points})-\text{(\# fitting parameters)}=20-4$.
We have found it to be $4.07\times 10^{-7}$ for the fit including $\g_1$.}  
The particle-like nature of the real-time conductivity remains extremely robust. 

We emphasize that the ``anomalous'' vortex-like responses described above (\rfig{im-frequ}) are not
consistent with the QMC data. Indeed, their small frequency conductivity (relative to $\s(\infty)$)  
is too small compared with the data. Further, the data shows an increasing second derivative as $w=0$ is 
approached, with no sign of an inflection point. 
 
\section{Proof of sum rules and asymptotics}    
\label{ap:sr}
We prove the conductivity sum rules: 
\begin{align}
  \int_0^\infty dw [\re\s(w)-\s(\infty)] &=0\,, \label{eq:sr1} \\
  \int_0^\infty dw [\re\hat\s(w)-\hat\s(\infty)] &=0\,, \label{eq:sr2} 
\end{align}
with $\hat\s= 1/\s$.
The sum rules were numerically verified to hold for the theory with Weyl coupling $\g$ in Ref.~\onlinecite{ws},
where the S-dual relation \req{sr2} was introduced. 
\req{sr1} was first discussed in Ref.~\onlinecite{sum-rules}, where it was 
proved to hold for a wide class of holographic theories. At first glance, the proof of Ref.~\onlinecite{sum-rules}, 
does not seem to apply to the actions under consideration in this work, except for the superconformal 
Yang-Mills fixed point. However, after a suitable transformation of the gauge equation of motion found in
Ref.~\onlinecite{sum-rules}, their proof can indeed be adapted to the theories we consider. (We are grateful 
to C.~Herzog for making this observation.) As discussed below, one assumption of Ref.~\onlinecite{sum-rules} needs to be
relaxed to cover our family of theories.
Below we present two independent proofs of the sum rules:  
the first one using a WKB analysis, and the second  
one adapted from Ref.~\onlinecite{sum-rules} to our theories. We note that currently only
the first approach captures the precise asymptotics, whereas the second only provides looser bounds. 

Ref.~\onlinecite{ws} in addition gave arguments for the validity of \req{sr1} for general CFTs, by relating the integral   
$\int_0^\infty dw\re\s(w)$ to an equal-time correlation function, which should not depend on temperature. 
\req{sr1} was further shown\cite{ws} to hold for the conductivity of the O($N$) CFT in the large-$N$
limit, as well as for free Dirac fermions. On the holographic side, Ref.~\onlinecite{ws2} 
extended the sum rules to finite momentum, and related them to Kramers-Kronig relations for the
retarded current-current correlators. The vanishing of the r.h.s.\ of \req{sr1} and \req{sr2} 
was then argued to be related to gauge invariance in the bulk. We here provide a definitive
proof on the holographic side, which holds for the entire family of theories considered in the main
body, and even beyond.  

The conductivity obtained from tree-level holography is a meromorphic function which only has poles in the lower half-plane
of complex frequency $w\in \mathbb C$. This latter property is a requirement from the retarded 
(causal) nature of the current-current correlation function out of which $\s$ is obtained. In the
previous section, we have seen that the bounds on the couplings ensure that no poles or zeros (QNMs)
appear for $\im w\geq 0$. If we can prove that the integrands decay sufficiently fast as $|w|\ra\infty$,
for $\im w\geq 0$, the sum rules follow from a simple contour integration in the upper half-plane. 

\subsection{WKB} 
We start with the equation of motion for the transverse gauge component $A_y$:
\begin{align}
  -\pd_z^2\psi+V_y(z)\psi = w^2\psi\,, \label{eq:psi-eq} 
\end{align}
where as before, $A_y=G_y(z)\psi(z)$, with $G_y=X_1^{-1/2}$. The $z$-variable is related to the
holographic coordinate $u$ by $dz/du=1/f$.
To study the conductivity,
we can set the momentum to zero, so that $V_y=V_{1,y}$, which is given in \req{V1y}. 
The conductivity is obtained via
\begin{align} \label{eq:sig_psi}
  \s(w)g_4^2=\frac{1}{iw}\frac{A_y'}{A_y}\Big|_{u=0}= \frac{1}{iw}\frac{\psi'(0)}{\psi(0)} \,, 
\end{align}
where in the last equality we have used the fact that $G_y'(u=0)=0$.  

\begin{figure}
\centering%
\includegraphics[scale=.59]{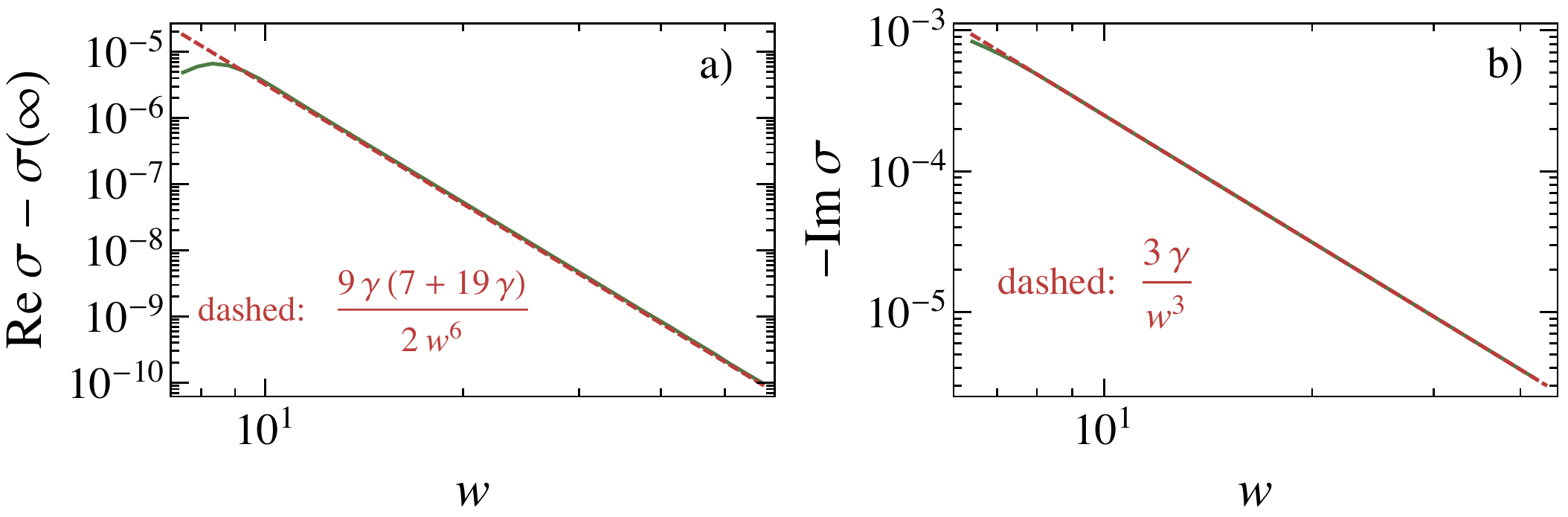}   
\caption{\label{fig:wkb} Log-log plots of the large-frequency conductivity, in units of $\s(\infty)$, for
$\g=1/12$ (solid green curves). The red dashed lines correspond to the WKB estimates, \req{sig_gam_wkb}. In (b), the absolute value of the
imaginary part is plotted.  
}  
\end{figure}

Let us start by assuming that $w$ lies on the real axis and $w>0$; we will analytically 
continue the results to the rest of the upper half-plane at the end. 
Since we are interested in the $w\gg 1$ limit, we can use the WKB approximation to solve \req{psi-eq}. 
We are looking for wave-like solutions with ``energy'' $w^2$ much greater than the
potential, the latter being in fact bounded. Let us rewrite the equation in a slightly different form: 
\begin{align}
  \pd_z^2\psi + w^2P^2\psi=0\,, \quad P=\sqrt{1-V_y(z)/w^2}\,.
\end{align}
We note that the function $P$ has no zeros in the entire range between the horizon and the UV boundary
($P\approx 1$ in the large $w$ limit), allowing us to safely use the WKB approximation. 
The latter states  
\begin{align}
  \psi &\to Q\exp\left[iw\int_0^zdz' P(z')\right]\,, \\
  &Q = Q_0 + w\inv Q_1 + w^{-2}Q_2+\cdots \\
 & \text{``$\to$'' means in the large-$w$ limit} 
\end{align}
We shall use the ``$\to$'' notation throughout. The first-order WKB approximation only keeps $Q_0$ (the zeroth
order approximation would set $Q=1$):  
\begin{align} \label{eq:Q0}
  Q_0 = \frac{1}{P^{1/2}} = \frac{1}{(1-V_y/w^2)^{1/4} }\,. 
\end{align}
At this order, the resulting conductivity is  
\begin{align}
  \s(w)g_4^2\to 1-i\frac{V_y'(0)}{4w^3}\,. \label{eq:wkb1} 
\end{align}
We recall that the potential is purely real, so that \req{wkb1} only tells us about the leading
$w\gg 1$ correction to the imaginary part of the conductivity, assuming $V_y'(0)\neq0$. In fact,
substituting the general form $X_1=\sum_{n=0}^\infty \Upsilon_1^{(n)}u^{3n}$ in the expression for
$V_{y}\big|_{q=0}=V_{1,y}$, \req{V1y}, we find 
\begin{align}
  V_{1,y}(0) &= 0 \,, & V_{1,y}'(0) &= 3\Upsilon_1^{(1)}\,, & V_{1,y}''(0)&= 0\,, \nn \\
  \pd_u^3V_{1,y}(0) &= 0 \,, & \pd_u^4V_{1,y}(0) &= -126\Upsilon_1^{(1)}(2+\Upsilon_1^{(1)})+360\Upsilon_1^{(2)} \,, & \pd_u^5V_{1,y}(0) &= 0\,,  \label{eq:V1y_exp}
\end{align}
with the next non-zero derivative being $\pd_u^7V_{1,y}(0)$, which involves a linear combination of $\Upsilon_1^{(1,2,3)}$.  
Generally, the form $X_1=\sum_{n=0}^\infty \Upsilon_1^{(n)}u^{3n}$ implies that only $\pd_u^nV_{1,y}(0)$
with $n\mod 3 =1$ will be finite. Now, taking the general Lagrangian for the gauge field, \req{L_gen}, 
we find that $\Upsilon_1^{(1)}=4\g$  
so that $\s(w)g_4^2\to 1-i\g/w^3$. It turns out we need to resort to at least a \emph{fourth-order} WKB approximation 
to capture the correct leading behavior of the real part. This means we need to determine the $Q_n$ up to
and including $n=3$. The $Q_n$ are determined recursively ($n\geq 1$): 
\begin{align} \label{eq:wkb-recursion}
  Q_n=\frac{i}{2}Q_0\int_0^z dz'\, Q_0(z')\pd_z^2Q_{n-1}(z')\,. 
\end{align}
(A simplification in the calculation is that $z\ra u$ as $u\ra 0$, so that one can forget 
about the distinction between the two variables $z$ and $u$ in that limit, which is relevant for the computation of $\s$.)  
It can then be shown that  
\begin{align}
  \s(w)g_4^2=\frac{1}{iw}\frac{\psi'(0)}{\psi(0)} \to 1+ \frac{Q_0'(0)}{iw}+\sum_{n=1} \frac{Q_{n-1}''(0)}{2w^{n+1}} \,. 
\end{align} 
From \req{wkb-recursion} and \req{Q0}, we see that $Q_n$ is real (imaginary) for $n$ even (odd). 
Including terms in the expansion up to $n=3$, we find
\begin{align}
  \re\s(w)g_4^2 &\to 1+ \frac{1}{32w^6}\left( -\pd_u^4V_{1,y}(0)+ 5[V_{1,y}'(0)]^2 \right) 
  - \frac{1}{w^{12}}\frac{1105[V_{1,y}'(0)]^4}{2048}\,, \label{eq:wkb_re} \\
  \im\s(w)g_4^2 &\to -\frac{V_{1,y}'(0)}{4w^3}+\frac{15[V_{1,y}'(0)]^3}{64w^9}\,. \label{eq:wkb_im} 
\end{align}
Let us take for example the theory with only the Maxwell-Weyl action, i.e.\ $\g\neq0$. 
Keeping the leading order contributions we find:
\begin{align} \label{eq:sig_gam_wkb}
  \s(w)g_4^2\to 1- i \frac{3\g}{w^3} +\frac{9\g}{2w^6}(7+19\g) \,.  
\end{align}
The term linear in $\g$ for the real part comes from $Q_3$ via $\pd_u^4V_{1,y}(0)$. It
is the dominant one, and this is the reason why we had to go to fourth order in the WKB expansion.
Indeed, the ratio of the linear to quadratic contributions (in absolute value) to the coefficient of $1/w^6$
is $7/(19|\g|)\geq 84/19\approx 4.4$, the latter value corresponding to $|\g|=1/12$.
\rfig{wkb} shows a comparison between the actual large-$w$ conductivity and the WKB approximation
for $\g=1/12$. We see that the agreement is excellent. We have further verified that the \emph{sign} of 
both the real and imaginary parts of $\s$ matches the WKB result for $\g>0$ and $<0$. This
proves that the integrands of \req{sr1} and \req{sr2} indeed decay sufficiently fast for the integrals
to converge.  The integrand also decays fast enough as $|w|\ra\infty$ in the upper half-plane.
In fact, the slowest decay occurs along the imaginary axis: letting $w\ra i\varpi$ in \req{sig_gam_wkb}  
yields:
\begin{align}
  \s(w\ra i\varpi)\to 1 + \frac{3\g}{\varpi^3} + \mc O(\varpi^{-6}) \,.  
\end{align} 
We have compared this result with the numerical solution for $\s(i\varpi)$ at various values of $\g$: 
the agreement is again excellent, as good as what was found on the real axis, \rfig{wkb}.     
We note that the imaginary time scaling can be directly compared with quantum Monte Carlo results\cite{ads-qmc,pollet}
on the O(2) QCP in 2+1D, for instance, but higher precision data is required to unambiguously identify the large-$\varpi$
scaling.  

We contrast the $1/w^6$ decay of $[\re\s-\s(\infty)]$ with the result for the QCP of the O($N$) model
in the large-$N$ limit\cite{damle}, see \rfig{comp}a,  
which obeys the slowest decay allowed by the sum rule, namely $1/w^2$.  

Yet different large-$\w$ scaling appears for a CFT of free Dirac fermions, namely
\begin{align}
  \re\s(\w/T)-\s(\infty)=-2\s(\infty)e^{-\omega/2T} +\mc O(e^{-\w/T})\,, \quad \w\gg T \,,
\end{align}
which follows from $\re\s/\s(\infty)=8\ln 2\;\de(\w/T)+\tanh(\w/4T)$, see for e.g.\ Refs.~\onlinecite{ssqhe} and \onlinecite{fritz08}.  
An exponential scaling also occurs for the free bosonic O($N$) model, also a CFT, for any $N\geq 2$. 
This can be simply seen by taking the leading order result of Ref.~\onlinecite{damle}
in $1/N$ and setting the interaction generated mass, $m$, to zero. This yields
$\re\s/\s(\infty)=1/\tanh(\w/4T)$ for $\w>0$, which, interestingly, is the inverse of the real part of the   
Dirac fermion conductivity given above.  
We emphasize that this exponentially fast approach to the asymptotic value $\s(\infty)$ found in these trivial  
CFTs is in contrast to the power law scalings found above.   
 
\subsection{Contraction maps}
We now provide a proof of the sum rules using a different approach, which involves
the use of contraction maps\cite{sum-rules}. 
We start with the general gauge action considered in Ref.~\onlinecite{sum-rules}:
\begin{align} \label{eq:eom_herzog}
  \mc A''+\frac{f'}{f}\mc A'+\left[ \frac{w^2}{f^2}-\frac{Y(u)}{u^2f} \right]\mc A=0\,. 
\end{align}
We have rewritten it using the variable $u$, instead of $r\propto 1/u$, which is used in that paper;
this explains the extra factor of $u^2$ in the coefficient of $\mc A$ in the above equation.
We have also set the function $\chi$ of Ref.~\onlinecite{sum-rules} to zero, because it is unecessary 
in our current analysis. The function $Y$ is frequency independent, and assumed to be non-negative\cite{sum-rules}. 
We note that \req{eom_herzog} does not have the same form as our equation of motion for $A_y$, \req{eom_Ay} 
(or see \req{eom_Ay_2}).  
However, one can perform a clever transformation, followed by a suitable choice of $Y$, to bring it into that form.  
One begins with the redefinition $\mc A=X_1^{1/2} A_y$. Then we choose 
\begin{align}\label{eq:Y}
  Y=\frac{u^2}{f}V_{1,y}\,,
\end{align}
where $V_{1,y}$ is precisely the potential term for the transverse gauge field found above, \req{V1y}. 
This leads to 
\begin{align} \label{eq:eom_Ay_2}
  A_y''+\frac{(fX_1)'}{fX_1} A_y'+\frac{w^2}{f^2}A_y=0\,,
\end{align}
which is the equation we were seeking, \req{eom_Ay}. These transformation are not surprising in light of
those that were performed in Appendix~\ref{ap:bounds} to obtain a Schr\"odinger type equation for $A_y$.
Indeed, the redefinition $A_y=X_1^{-1/2}\psi$ was performed, which corresponds to what we have done above.
We can thus relate $\mc A=\psi$. It follows that the expression for the conductivity is the same as above, \req{sig_psi}:
\begin{align} \label{eq:sig_psi}
  \s(w)g_4^2= \frac{1}{iw}\frac{\psi'(0)}{\psi(0)} \,, 
\end{align}
where we now use $\psi$ to denote $\mc A$. Ref.~\onlinecite{sum-rules} then transformed the 2nd order linear 
ordinary differential equation (DE) into a 1st order non-linear DE. To obtain the asymptotic behavior of the
conductivity, they constructed a contraction map to identify the large-$w$ behavior. 
The details are somewhat involved, so we refer the interested reader to Ref.~\onlinecite{sum-rules}.  
The result is 
\begin{align} \label{eq:s_fn}
  \s(w)g_4^2= \frac{1-s(w)}{1+s(w)}\,; \qquad  s(w) \sim w^{-2\Delta}\; \text{ as }\; w\ra \infty \,.  
\end{align}
The exponent $\Delta$ appearing in the function $s(w)$ 
is obtained from the near-boundary scaling of $Y$: $Y\sim u^{2\Delta}$  
as $u\ra 0$. An assumption in Ref.~\onlinecite{sum-rules} is that $Y(u)\geq 0$. As seen in Ref.~\onlinecite{myers11} 
and in Appendix~\ref{ap:bounds}, this is not satisfied by the family of theories that we consider (see \rfig{V1}).
However, if we restrict ourselves to theories that do not possess very deep negative potential wells, 
no instabilities occur and all the QNMs remain in the lower half-plane (see Appendix~\ref{ap:bounds}).
As this latter requirement seemed to be the reason why the authors of Ref.~\onlinecite{sum-rules}  
restricted themselves to $Y\geq0$, we expect that relaxing this 
condition as we just described will not change the validity of the proof found in Ref.~\onlinecite{sum-rules}. 
Using \req{Y} and the expansion for $V_{1,y}(u)$, \req{V1y_exp}, we see that as $u\ra 0$  
\begin{align} 
  Y\approx \frac{u^2}{1-u^3}[uV_{1,y}'(0) + u^4\pd_u^4V_{1,y}(0)+\cdots] 
  \approx u^3 V_{1,y}'(0)\,,   
\end{align}
so that 
\begin{align} 
  2\Delta=3=D, 
\end{align}
when $V_{1,y}'(0)\neq 0$. We have emphasized that the power corresponds to the
dimension of the spacetime of the boundary CFT, i.e.\ $D=2+1$. This is
the same scaling with $D$ as that of the ``pressure contribution'' of Ref.~\onlinecite{sum-rules},
which in fact vanishes in $D=2+1$ but not in $D=3+1$.  
Now, we recall that $\s(w)^*=\s(-w)$, so that the function $s(w)$ must
satisfy the same condition, $s(w)^*=s(-w)$.  
Given the power law $s(w)\sim w^3$, the function must thus be purely
imaginary: $s(w)=iaw^3$, $a\in\mathbb R$, in the large-$w$ limit. We thus find the asymptotic
expansion:
\begin{align}
  \s(w)g_4^2=(1-2a^2w^{-6}+2a^4w^{-12}+\cdots)+i(-2aw^{-3}+2a^3w^{-9}+\cdots)\,. 
\end{align} 
The $w$-dependences of the real and imaginary parts are in agreement with what was found using the WKB
expansion, \req{wkb_re} and \req{wkb_im}, respectively. Indeed, the
imaginary part decays like $1/w^3=1/w^D$ while the real part goes like $\re \s-\s(\infty)\sim 1/w^6=1/w^{2D}$.  
We further note that the subleading corrections, which take the form $1/w^{6+6n}$ and $1/w^{3+6n}$ ($n\geq1$),
for the real and imaginary parts match the WKB analysis. 

What happens if $V_{1,y}'(0)=0$, i.e.\ $\g=0$? In that case, we must consider the 
next term in the series expansion of $Y(u)$: $Y(u)\sim u^6\pd_u^4V_{1,y}(0)$, where
$\pd_u^4V_{1,y}(0)\propto\Upsilon_1^{(2)}$ (see \req{V1y_exp}). This leads to $2\Delta=6$ 
instead of $2\Delta =3$ as found above. 
The condition $\s(-w)=\s(w)^*$ now requires $s(w)$ to be a real function:
$s(w)=\t a w^{-6}$ in the large-$w$ limit, where $\t a\in\mathbb R$. Using \req{s_fn}, we thus again find that
$\re\s-\s(\infty)\sim 1/w^6=1/w^{2D}$ as above. This agrees with the WKB analysis.  
Both the latter and the contraction map methods do not provide the scaling for the imaginary part
in this case, at least at the expansion level considered here. The numerical results at $\g=0$ for the imaginary 
part do not provide a clear asymptotic scaling in the large-$w$ limit, unlike when $\g=0$. 
One difference is that $\im\s$ oscillates more (changing sign in the process) compared with the
$\g\neq0$ case. We leave this for further investigation. 

\section{General self-duality}
Generally, under S-duality, we have $\hat X_I^J=-\varepsilon_I^{\;\; B}(X\inv)_B^C\,\varepsilon_C^{\si J}$,
where $\varepsilon_A^{\si B}$ is the matrix representation of the fully anti-symmetric tensor $\varepsilon_{ab}^{\si cd}$,
with $\varepsilon_{txyu}=\sqrt{-g}$.  It is given by the anti-diagonal matrix 
$(\varepsilon_1^{\;\;6},\varepsilon_2^{\;\;5},
\varepsilon_3^{\;\;4},\varepsilon_4^{\;\;3},\varepsilon_5^{\;\;2},\varepsilon_6^{\;\;1})=(f,-f,1,-1,f\inv,-f\inv)$.
It is traceless, has unit determinant and $\varepsilon^2=-1$. We thus have\cite{myers11}
\begin{align}  
  \hat X_I^{J}=\diag(\hat X_1,\hat X_1,\hat X_3, \hat X_4, \hat X_5,\hat X_5) 
=\diag(X_5\inv,X_5\inv,X_4\inv,X_3\inv,X_1\inv,X_1\inv)\,,  
\end{align}
where we have used rotational symmetry to set $X_1=X_2$ and $X_5=X_6$. For the subspace of terms considered
in the main body, in addition we had $X_1=X_5$ and $X_3=X_4$. In that case, a self-dual conductivity followed 
simply from the requirement $X_1= 1$. It was noted that this condition can be satisfied for theories with an
arbitrary number of derivatives. 
We here describe the more general situation, allowing for $X_1\neq X_5$, in which case the general condition to obtain
a self-dual conductivity is  
\begin{align} \label{eq:gen_self-dual}
  X_5=1/X_1 \implies \s(w)=\s(\infty)\;\; \forall w\,.
\end{align}
Note that this condition encompasses the case discussed in the main body, namely $X_1=X_5$, 
leading to $X_1=1$. 
Using \req{gen_self-dual} one can rewrite the $q=0$ equation for $A_y$ as follows:
\begin{align} \label{eq:gen_sd}
  A_y''+\frac{(X_5f)'}{X_5f}A_y'+\frac{w^2}{(X_5f)^2}A_y=0\,. 
\end{align}
This can be mapped to a harmonic equation via the change of variables $dz/du=1/(X_5f)$:
$\pd_z^2A_y+w^2A_y=0$. This is exactly analogous to the case\cite{m2cft} where $X_1=X_5=1$, but with   
$f$ replaced by $X_5f$. The solution for $A_y$ that is in-falling at the horizon is 
$A_y=e^{iwz}=\exp[iw\int_0^u du'/(X_5(u')f(u'))]$.
Using \req{sig}, this leads to a frequency-independent conductivity 
as long as $f(0)X_5(0)$ is finite at the UV boundary $u=0$. This is a reasonable
condition; for example, it is satisfied by all the theories described in the main text.  

It will be interesting to investigate the existence of such theories in holographic actions with
other types of HD terms. For example, Ref.~\onlinecite{sum-rules} has considered a gauge
field propagating in a general background spacetime with line-element: 
\begin{align}
  ds^2=-\frac{r^2}{L^2}\t f(r)e^{-\chi(r)}dt^2+\frac{r^2}{L^2}(dx^2+dy^2)+\frac{L^2\,dr^2}{r^2\t f(r)}\,,  
\end{align}
where we are using $r\propto 1/u$ as the radial coordinate. We have added a tilde to $f$ to distinguish 
it from the one used in the present work. The assumptions\cite{sum-rules} on the spacetime are quite general: 
first, it contains a horizon at $r=r_h$, where $\t f(r_h)=0$ and $\t f'(r_h)\neq 0$; second, it asymptotes to 
Anti-de Sitter with a radius of curvature $L$ (which we now set to 1) at large $r$: $\t f=1+\cdots$ and $e^{-\chi}f=1+\cdots$.
The equation of motion for the gauge field dual to the CFT current $J_\mu$  was chosen to be (for $\mu=y$):    
\begin{align} \label{eq:eom_herzog_2}
  \pd_r^2A_y + \frac{\pd_rF}{F}\pd_rA_y+\left[ \frac{w^2}{F^2} - \frac{Y}{F} \right]A_y =0\,,
\end{align} 
where $F=r^2\t f e^{-\chi/2}$. We note that this reduces to \req{eom_herzog} upon changing variables to $u\propto 1/r$,
setting $\chi=0$ and $\t f=1-r_h^3/r^3=f$ (i.e.\/ the same spacetime we have considered in the main text).    
If we set $Y=0$ in \req{eom_herzog_2}, it takes the same form as \req{gen_sd}.  
In fact, changing variables back to $u=r_h/r$, we obtain:
\begin{align}
  A_y''+ \frac{(\t f e^{-\chi/2})'}{\t f e^{-\chi/2}} A_y' + \frac{w^2}{(\t f e^{-\chi/2})^2}A_y =0\,,
\end{align}
which maps to \req{gen_sd} when $fX_5\leftrightarrow \t fe^{-\chi/2}$. The same conclusion thus holds: this general  
family of holographic theories has a self-dual, frequency-independent conductivity. We note that this 
is of relevance to theories with purely gravitational HD terms, such as those considered in Ref.~\onlinecite{bai}.   

\section{Solving the general equation of motion}
We write 
\begin{align}
  A_y=(1-u)^bF(u)\, , \quad b=-iw/3=-i\w/4\pi T\,,
\end{align}
where $b$ was chosen such that $F$ is regular at the horizon, $u=1$;
we are working with waves that are in-falling into the black brane. 
We then get the general equation for $F$ at zero momentum, $q=0$: 
\begin{align}
  F''+\left[\frac{(X_1f)'}{X_1f} -\frac{2b}{1-u} \right]F' 
  + \left[ \frac{b(b-1)}{(1-u)^2} -\frac{b}{1-u}\frac{(X_1f)'}{X_1f} + \frac{w^2}{f^2} \right]F=0\,;
\end{align}
with boundary conditions:
\begin{align}
  F(1) &=1\,, \\
  F'(1) &=-b \left[ 1+\frac{1}{1+2b}\frac{X_1'(1)}{X_1(1)} \right]  \,.
\end{align}

\section{Diffusion constant and susceptibility}
\label{ap:dx}
\begin{figure}
\centering%
\includegraphics[scale=.33]{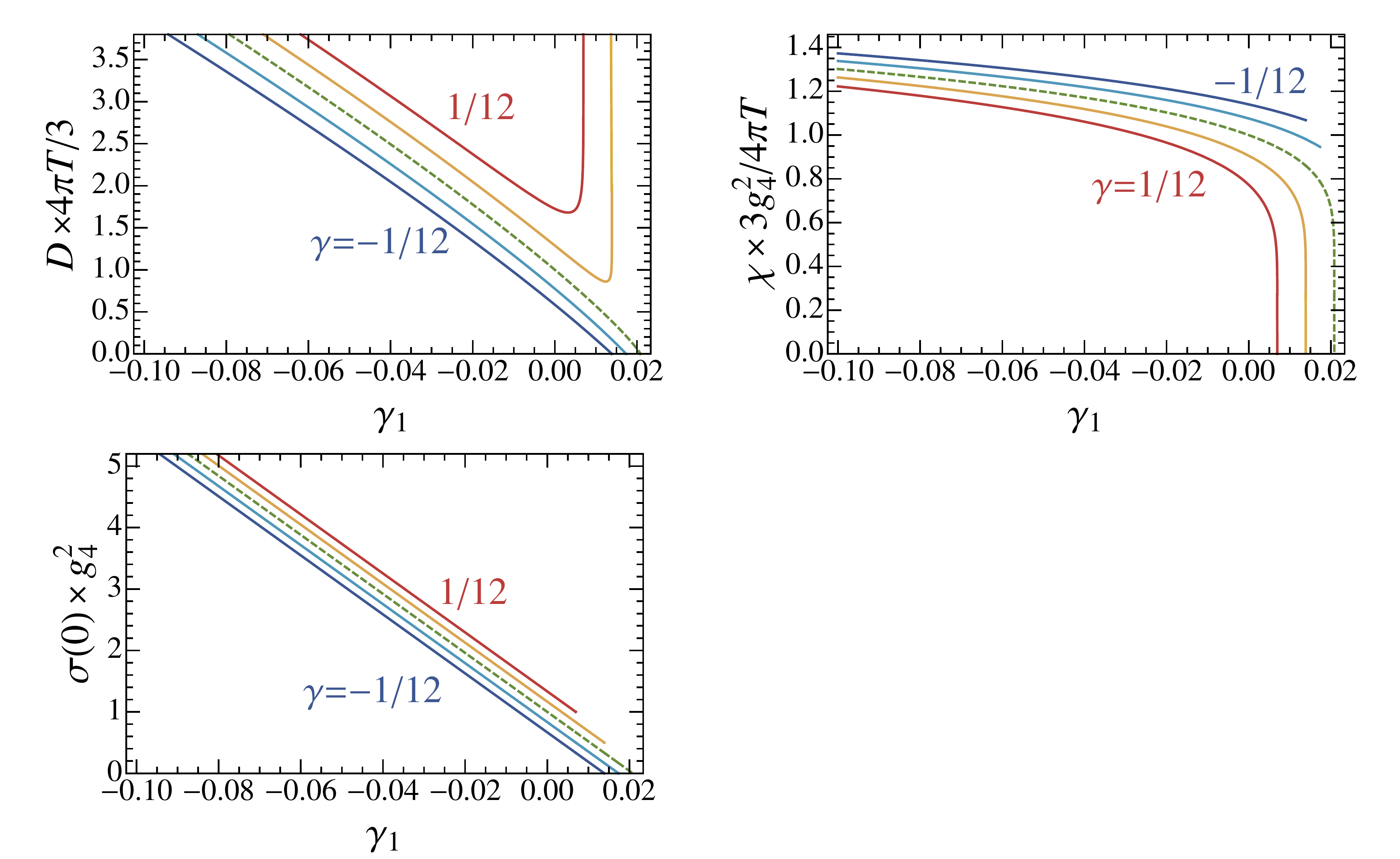}      
\caption{\label{fig:dxs} 
Plot of the diffusion constant $D$, charge susceptibility $\chi$, and d.c.\ conductivity $\s(0)$ as a function of $\g_1$,
for $\g=\pm 1/12,\pm 1/24, 0$. The $\g$ dependence is monotonic. The dashed lines correspond to
$\g=0$. 
}  
\end{figure}

We here describe the computation of the diffusion constant $D$ and charge susceptibility $\chi$.
We note that they obey the Einstein relation, $\s(0)=\chi D$. Generalizing the result
of Ref.~\onlinecite{myers11}, the diffusion constant can be shown to be:
\begin{align}
  D= \frac{3}{4\pi T} \s(0)g_4^2 \int_0^1 \frac{du}{X_3(u)}\,,
\end{align}
where
\begin{align}
  \s(0)g_4^2=X_1(1)\,.
\end{align}
We examine the case where only $\g_1\neq0$: $X_3(u)=1-48\g_1 u^6$, and $\g_1<1/48$. 
We find that 
\begin{align}
  D\approx 
  \begin{cases}
    c_1 |g_1|^{5/6}\,, & \g_1 \ra -\infty \\
    (1-48\g_1)\left[ -\tfrac{1}{6}\ln(1-48\g_1)+ c_2\right]\,, & \g_1\ra 1/48 
  \end{cases}
\end{align}
where $c_1=8(4/3)^{1/6}\pi$ and $c_2=(\sqrt 3 \pi+\ln 432)/12$. This leads to a divergence of the
diffusion constant as $\g_1\ra-\infty$, and its vanishing as $\g_1\ra 1/48$. 
Using the Einstein relation we obtain the charge susceptibility:
\begin{align}
  \chi\sim
  \begin{cases}
    |g_1|^{1/6}\,, & \g_1 \ra -\infty \\
    \frac{-1}{\ln(1-48\g_1)}\,, & \g_1\ra 1/48 
  \end{cases}
\end{align}
Thus, $\chi$ shows the same divergence/vanishing as $D$ in the asymptotic regions; this is summarized in \rfig{map}b.
However, this needs not be the case in general. \rfig{dxs} illustrates the dependence of $D,\chi,\s(0)$
on $\g_1$ for different values of $\g$. We note that when $\g=1/12,1/24$, $D$ diverges while $\chi$ vanishes as 
$\g_1\ra 1/144, 1/72$, respectively. In both cases the d.c.\ conductivity remains finite in that limit, as shown in the lower panel of \rfig{dxs}. 
\pagebreak 

%

\end{document}